\documentclass[11pt, oneside]{article}
\usepackage{geometry}
\usepackage{setspace} 
\usepackage{lineno} 
\usepackage[parfill]{parskip}
\usepackage{amssymb}
\usepackage{amsmath}
\usepackage{dsfont}
\usepackage{mathptmx} 
\usepackage{graphicx}
\usepackage[labelfont=bf]{caption} 
\usepackage{float} 
\usepackage[
	style=authoryear,
	citestyle=authoryear,
	maxbibnames=6,
	giveninits=true,
	backend=bibtex,
	url=false,
	doi=false,
	isbn=false,
	eprint=false
	]{biblatex}
\AtEveryBibitem{\clearfield{note}}
\newcommand{\beginsupplement}{%
        \setcounter{table}{0}
        \renewcommand{\thetable}{S\arabic{table}}%
        \setcounter{figure}{0}
        \renewcommand{\thefigure}{S\arabic{figure}}%
     }

\title{Fast fitting of neural ordinary differential equations by Bayesian neural gradient matching to infer ecological interactions from time series data}
\author{Willem Bonnaff\'e$^{1,2}$ \& Tim Coulson$^2$}
\date{}

\begin{document}
\maketitle
\vspace{-0.5cm}
\pagenumbering{gobble}

1. Big Data Institute, University of Oxford, Old Road Campus, Oxford OX3 7LF 

2. Department of Biology, University of Oxford, Zoology Research and Administration Building, 11a Mansfield Road, Oxford OX1 3SZ

\begin{center}
\textbf{Abstract}
\end{center}

1. Inferring ecological interactions is hard because we often lack suitable parametric representations to portray them.
Neural ordinary differential equations (NODEs) provide a way of estimating interactions nonparametrically from time series data. %, with no assumptions required for the functional form of the interactions.
NODEs, however, are slow to fit, and inferred interactions have not been truthed.
2. We provide a fast NODE fitting method, Bayesian neural gradient matching (BNGM), which relies on interpolating time series with neural networks, and fitting NODEs to the interpolated dynamics with Bayesian regularisation.
We test the accuracy of the approach by inferring ecological interactions in time series generated by an ODE model with known interactions. 
We also infer interactions in experimentally replicated time series of a microcosm featuring an algae, flagellate, and rotifer population, as well as in the hare and lynx system.
3. Our BNGM approach allows us to cut down the fitting time of NODE systems to only a few seconds.
The method provides accurate estimates of ecological interactions in the artificial system, as linear and nonlinear true interactions are estimated with minimal error.
In the real systems, dynamics are driven by a mixture of linear and nonlinear ecological interactions, of which only the strongest are consistent across replicates.
4. Overall, NODEs alleviate the need for a mechanistic understanding of interactions, and BNGM alleviates the heavy computational cost. 
This is a crucial step availing quick NODE fitting, cross-validation, and uncertainty quantification, as well as more objective estimation of interactions, and complex context dependence, than parametric models. 

\vspace{0.5cm}


\textbf{Emails:}
willem.bonnaffe@nds.ox.ac.uk;
tim.coulson@zoo.ox.ac.uk;

\newpage
\pagenumbering{arabic}
\setcounter{page}{1}
% \linenumbers
\setstretch{1.5}

\section{Introduction}

The concept of population is central in ecology (\cite{Berryman2002}).
Ecologists have had a longstanding interest in finding laws that govern population dynamics, namely changes in the number of individuals in the populations (\cite{Lawton1999,Turchin1999a}).
Population dynamics can be characterised by a logistic growth, or similar forms, limited by ecological interactions with other organisms, and by the state of the environment (\cite{Turchin2001c,Berryman2003}).
Intra-specific interactions correspond to interactions between individuals of different sex, age or size classes, belonging to the same species (\cite{Turchin2001c}).
Inter-specific interactions are interactions between individuals from different species, be it competitors, preys, predators, or pathogens (\cite{Turchin2001c,Berryman2003}).
These interactions can cause populations to have lagged effects impacting their own growth, often called feedback effects, mediated by their impact on the other populations they interact with (\cite{Berryman1997}).

Characterising these interactions has been a longtime challenge. 
Ecologists started analysing time series data with parametric models (\cite{Royama1984,Kendall1999,Ives2003,Gross2005}), as time series of population counts are the most commonly collected long-term data in biology (\cite{Kendall1999}).
Initial analysis involved fitting simple auto-regressive linear models to time series of a single species, leading to contentious interpretations of interactions thereby inferred (e.g. \cite{Berryman1997}). 
For instance, Royama et al. interpreted higher order lags as evidence of species interactions (\cite{Royama1984}), while Lande et al. interpreted them as age-structure signatures (\cite{Lande2002}). 
Coulson et al. showed they can even be caused by interactions between the sexes (\cite{Mysterud2002}).
Jonzen et al. added doubt over interpreting lags by demonstrating that autocorrelation in environmental noise could prevent altogether the reliable estimation of lag effects in single species time series data (\cite{Jonzen2002}).
More recent work has investigated time series of multiple species, environmental factors, and has mechanistically modelled various ecological interactions (e.g. \cite{Bruijning2019,Rosenbaum2019,Adams2020}).
In these models, ecological interactions are quantified explicitly by specific parameters, rather than phenomenologically with lags.
This allowed for a more thorough quantification of interactions and comparison of alternative ecological interactions architectures.

However, ecologists still face two main obstacles when estimating ecological interactions from time series data.
The first is that interactions are highly context-dependent, so that they change in time with the state of the ecosystem and of the environment (\cite{Song2020}).
Ecological interactions were traditionally considered linear or fixed, yet there is substantial evidence that this is not the case in nature (e.g. \cite{Bonsall2003,Gross2005,Kendall2005,Ushio2018,Bruijning2019,Rosenbaum2019,Bonnaffe2021b}).
The effect of the population on itself depends on the density of individuals (e.g. \cite{Lingjaerde2001, Moe2005, Brook2006}); while predation rates can depend on the density of the predator (\cite{Jost2000,Yoshida2003}).
Many vital rates underpinning ecological interactions are age- and size-dependent (\cite{Bonnaffe2018,Bonnaffe2021b}), and governed by environmental variables, such as temperature (\cite{Brown2004a}).
Interactions also change following evolution of the traits that underpin them (\cite{Turchin2003,Yoshida2003}).
This makes it virtually impossible to model the full complexity of ecological interactions (\cite{Lawton1999,Kendall1999}).

This leads to the second obstacle, known as structural sensitivity, namely sensitivity of the results to the structure of the model (\cite{Wood2001,Adamson2013}).
Because of the complexity of the interactions, we often lack suitable mathematical representations to portray them (\cite{Jost2000,Wood2001,Ellner2002,Wu2005}).
Parametric representations of the interactions are assumed \textit{a priori}, which means that any interaction quantified is ultimately contingent on this arbitrary choice, and hence potentially biased (\cite{Jost2000,Wood2001,Ellner2002,Wu2005}).
Parametric inference of ecological interactions from time series data therefore only provides qualitative evidence, requiring further experimental verification and quantification (\cite{Kendall1999}). 

Nonparametric modelling provides a powerful alternative that can help solve these problems (e.g. \cite{Jost2000,Wood2001,Ellner2002,Wu2005,Pasquali2018}).
Nonparametric forms give more freedom to researchers wishing to  model population dynamics, and allow a test of whether the linear or linearised assumption of standard models is warranted.
Interactions are quantified as the sensitivity of the nonparametric approximation of the dynamics with respect to other state variables (\cite{Sugihara2012,Ushio2018}).
Nonparametric models require minimal assumptions regarding the mathematical nature of ecological interactions (\cite{Jost2000,Gross2005}), and hence provide interaction estimates that are more robust to model structure (\cite{Wood2001}).
In particular, artificial neural networks (ANNs) offer a promising, yet underused, nonparametric alternative to linear functional forms.
In previous work, we introduced a powerful framework, relying on neural ordinary differential equations (NODEs, \cite{Chen2018}) to approximate the dynamics of populations nonparametrically, from which we derive ecological interactions (\cite{Bonnaffe2021a}).
More specifically, the ANNs embedded in the ODEs learn nonparametrically the shape of the per capita growth rate of the populations and its dependence on the state variables of the system (\cite{Bonnaffe2021a}).
Combined with the Geber method (\cite{Hairston2005}), we are able to estimate the direction, strength, and degree of nonlinearity of interactions.

One limitation of the approach lies in the computational cost of fitting the NODEs (\cite{Chen2018,Bonnaffe2021a}).
This is due to the fact that NODEs, as with ODEs, need to be simulated over the entire range of the time series in order to compute the likelihood of the trajectories of the model.
This can be avoided by using gradient matching, which requires interpolating the time series, and fitting the ODEs directly to the interpolated dynamics (\cite{Jost2000,Aarts2001,Ellner2002}).
Although a similar approach has been proposed (see \cite{Treven2021}), there are no implementations of it to fitting NODEs, in spite of its great potential for cutting down computational costs.
In addition, given the novelty of the framework, the accuracy and robustness of NODEs in estimating ecological interactions remain largely unexplored.
Most of the work to date is concerned with the accuracy of the fitted trajectories and of the forecasts (\cite{Mai2016,Treven2021,Frank2022}), while little attention has been given to the functional form of the processes that are producing the dynamics approximated by NODEs (but see \cite{Hu2020} for a step in this direction).
It is important to understand to what extent the neural networks embedded within NODEs carry meaningful biological information (\cite{Novak2021}).

In this manuscript, we first introduce a novel fitting technique for NODEs, Bayesian neural gradient matching (BNGM). 
The method extends gradient matching by using neural networks to interpolate the time series data instead of splines (\cite{Ellner2002}), and Bayesian regularisation to fit NODEs to the interpolated dynamics (\cite{Cawley2007}). 
This cuts down the fitting time of NODEs to only a few seconds, compared to about 30 minutes in our previous work (\cite{Bonnaffe2021a}), allowing for efficient cross-validation, and uncertainty quantification.
We then demonstrate that NODEs are highly accurate in recovering ecological interactions in an artificial three-species prey-predator system where truth is known.
Finally, we conclude the work by characterising ecological interactions in three replicates of an experimental three-species prey-predator system with an algae, flagellate, and rotifer (\cite{Hiltunen2013}), as well as in the classic hare and lynx time series (\cite{Odum1972}).
We find that only main interactions, between the algae and the rotifer, are conserved across the three replicates, and not the interactions of the flagellate with the other species.
We also find that in most cases linear interactions are sufficient to explain the dynamics apart from nonlinearity in the effect of the prey on the top predator in both the rotifer and lynx.

\section{Material and Methods}

\subsection{Method overview}

We provide a nonparametric method for estimating ecological interactions from time series data of species density. 
We do this by approximating the dynamics of each species with neural ordinary differential equations (NODEs, \cite{Bonnaffe2021a}). 
We then compute ecological interactions as the sensitivity of these dynamics to a change in the respective species densities (\cite{Sugihara2012,Bonnaffe2021a}).
We provide a novel method, Bayesian neural gradient matching (BNGM), allowing us to fit NODE systems in a only a few seconds.

\subsection{Neural ordinary differential equation}

A NODE is a class of ordinary differential equation (ODE) that is partly or entirely defined as an artificial neural network (ANN) (\cite{Chen2018}).
They are useful to infer dynamical processes nonparametrically from time series data (\cite{Bonnaffe2021a}).
We choose NODEs over standard statistical approaches because they offer two advantages. 
The first is that NODEs approximate the dynamics of populations nonparametrically.
NODEs are therefore not subjected to incorrect model specifications (\cite{Jost2000,Adamson2013}).
This provides a more objective estimation of the inter-dependences between state variables. 
The second advantage is that it is a dynamical systems approach. 
So that the approach includes lag effects through interacting state variables, not only direct effects between them. 

We first consider a general NODE system,

\vspace{-0.5cm}
\begin{equation}
    \frac{dy_i}{dt} = f_p \left(y,\theta_i \right), \\
\end{equation}

where $dy_i/dt$ denotes the temporal change in the $i^{th}$ variable of the system, $y_i$, as a function of the other state variables $y = \{ y_1, y_2, ..., y_I\}$.
The function $f_p$ is a nonparametric function of the state variables and its shape is controlled by the parameter vector $\theta_i$.
In the context of NODEs, $f_p$ is an ANN.
The most common class of ANN used in NODEs are single-layer fully connected feedforward ANNs (e.g. \cite{Wu2005}), also referred to by single layer perceptrons (SLPs, e.g. \cite{Bonnaffe2021a}),

\vspace{-0.5cm}
\begin{equation}
    f_p \left(y, \theta_i \right) = f_\lambda \left( \theta_i^{(0)} + \sum_{j=1}^{J} \theta^{(1)}_{ij} f_\sigma \left( \theta^{(2)}_{ij} + \sum_{k=1}^{K} \theta^{(3)}_{ijk} y_k \right) \right),
\end{equation}

which feature a single layer, containing $J$ neurons, that maps the inputs, here the state variables $y$, to a single output, the dynamics of state variable $i$, $dy_i/dt$.
The parameter vector $\theta_i$ contains the weights $\theta^{(l)}$ of the connections in the SLPs.
SLPs can be viewed as weighted sums of activation functions $f_\sigma$, which are usually chosen to be sigmoid functions $f(x) = 1/(1+\exp(-x))$.
The link function $f_\lambda$ allows to map the output of the network to a specific domain, for instance applying tanh will constrain the dynamics between -1 and 1, $dy_i/dt \in~]-1,1[$. 

This general form can be changed to represent biological constraints on the state variables.
In particular for population dynamics, the state variables are strictly positive population densities, $y_i = N_i \in~\mathcal{R^+}$.
We could hence re-write equation (1) as, $dN_i/dt = f_p(N,\theta_i)N_i$, where the SLPs approximate the per-capita growth rate of the populations.
More details regarding these models can be found in our previous work (\cite{Bonnaffe2021a}).

\subsection{Fitting NODEs by Bayesian neural gradient matching}

In this section, we describe how to estimate the parameters $\theta$ of the NODE system given a set of time series. 
Fitting NODEs can be highly computationally intensive, which hinders uncertainty quantification, cross-validation, and model selection (\cite{Bonnaffe2021a}).
We solve this issue by introducing BNGM, a computationally efficient approach to fit NODEs.
The approach involves two steps (Fig. 1).
First, we interpolate the state variables and their dynamics with neural networks (Fig. 1, red boxes).
Second, we train each NODE to satisfy the interpolated state and dynamics (Fig. 1, blue boxes).
This bypasses the costly numerical integration of the NODE system and provides a fully mathematically tractable expression for the posterior distribution of the parameter vector $\theta$. 
We coin the term BNGM to emphasise two important refinements of the standard gradient matching algorithm (\cite{Ellner2002}). 
The first is that we use neural networks as interpolation functions, and the second is that we use Bayesian regularisation to limit overfitting and estimate uncertainty around parameters (\cite{Cawley2007}).

%% figure
\newpage
\begin{figure}[H]
\includegraphics[width=\linewidth,page=1]{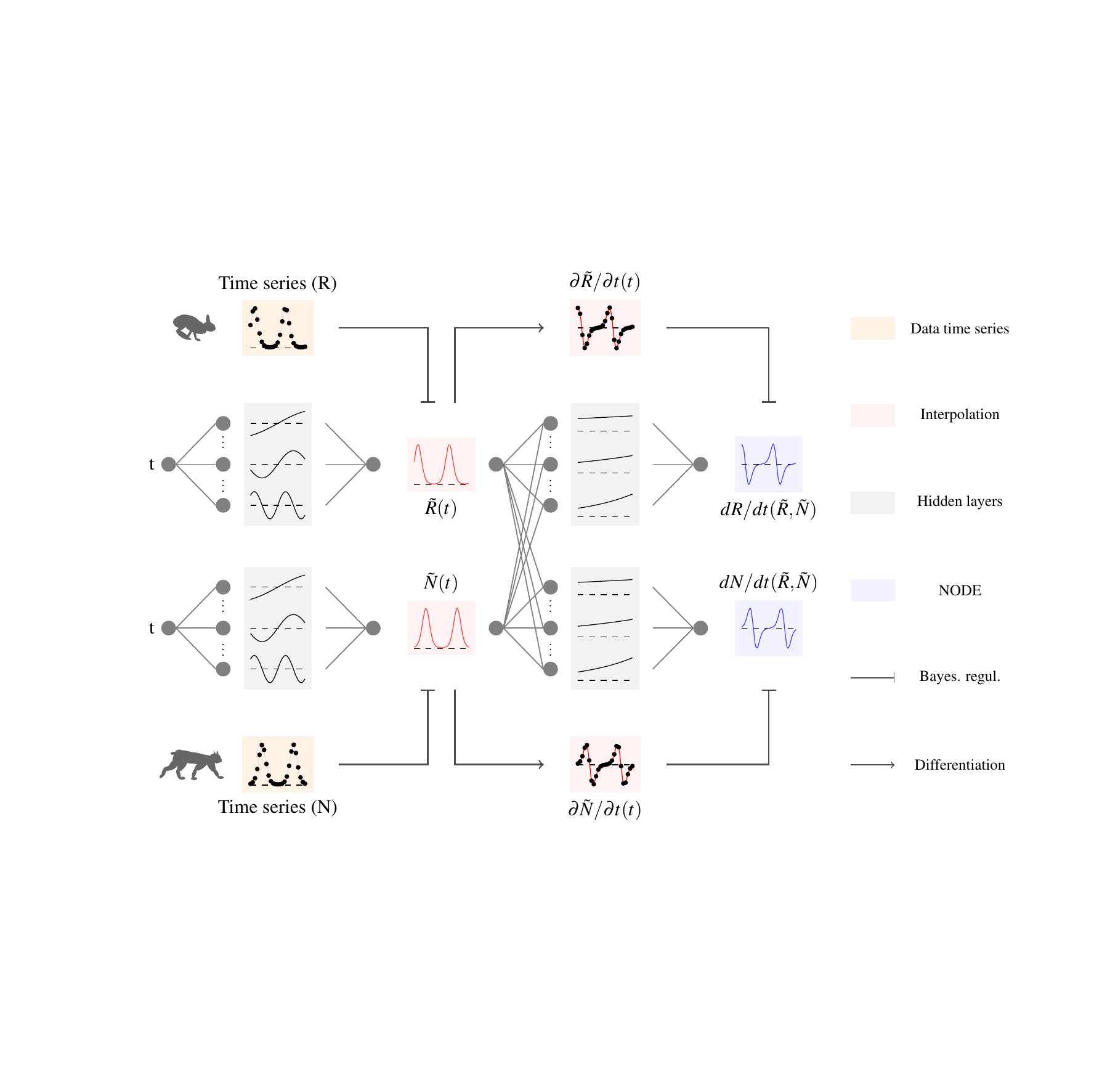}
\vspace{-3cm}
\caption{
    \textbf{Overview of fitting neural ordinary differential equations (NODE) by Bayesian neural gradient matching (BNGM).}
    In a first step we compute a continuous time approximation (interpolation) of each state variables, here the prey $\tilde{R}(t)$ and predator density $\tilde{N}(t)$ (red boxes).
    To do that we fit an ANN, that takes time as input, to each time series, via Bayesian regularisation.
    Interpolated dynamics of populations can then be computed by taking the derivative of the ANN with respect to time, $\partial\tilde{R}/\partial t$ and $\partial\tilde{N}/\partial t$.
    In a second step, we fit each NODE, $dR/dt$ and $dN/dt$ (blue boxes), to the interpolated dynamics.
    To do that we fit an ANN, which takes as input the interpolated variables $\tilde{R}(t)$ and $\tilde{N}(t)$, to the interpolated dynamics $\partial\tilde{R}/\partial t$ and $\partial\tilde{N}/\partial t$, via Bayesian regularisation.
}
\end{figure}
\newpage

\textbf{Interpolating the time series}

The first step is to interpolate the time series and differentiate it with respect to time in order to approximate the state and dynamics of the variables.
We perform the interpolation via nonparametric regression of the interpolating functions on the time series data,

\vspace{-0.5cm}
\begin{equation}
    Y_{it} = \tilde{y}_i(t,\omega_i) + \epsilon^{(o)}_{it},
\end{equation}

where $Y_{it}$ is observed value of the state variable $i$ at time $t$, $\tilde{y}_i(t,\omega_i)$ is the value predicted by the interpolation function given the parameter vector $\omega_i$, and $\epsilon^{(o)}_{it}$ is the observation error between the observation and prediction. 
The interpolation function is chosen to be a neural network,

\vspace{-0.5cm}
\begin{equation}
    \tilde{y}_i (t,\omega_i) = f_\lambda \left( \omega_i^{(0)} + \sum_{j=1}^{J} \omega^{(1)}_{ij} f_\sigma \left( \omega^{(2)}_{ij} + \omega^{(3)}_{ij} t \right) \right),
\end{equation}

where the parameter vector $\omega_i$ contains the weights $\omega^{(l)}$ of the network.
We can further differentiate this expression with respect to time to obtain an interpolation of the dynamics of the state variables (Fig. 1, red boxes), 

\vspace{-0.5cm}
\begin{equation}
    \frac{\partial \tilde{y}_i}{\partial t} (t, \omega_i) = \sum_{j=1}^{J} \omega^{(1)}_{ij} \omega^{(3)}_{ij} \frac{\partial f_\sigma}{\partial t} \left(\omega^{(2)}_{ij} + \omega^{(3)}_{ij} t \right) \frac{\partial f_\lambda}{\partial t} \left ( \omega^{(0)}_{i} + \sum_{k=1}^{J} \omega^{(1)}_{ik} f_\sigma \left( \omega^{(2)}_{ik} + \omega^{(3)}_{ik} t \right) \right ). 
\end{equation}

\textbf{Fitting NODEs to the interpolated time series}

The second step is to train the NODE system (Eq. 1) to satisfy the interpolated dynamics.
Thanks to the interpolation step, this simply amounts to performing a nonparametric regression of each NODE (Eq. 1) on the interpolated dynamics (Eq. 5),

\vspace{-0.5cm}
\begin{equation}
    \frac{\partial \tilde{y}_i}{\partial t} (t, \omega_i) = \frac{dy_i}{dt} \left( \tilde{y},\theta_i \right) + \epsilon^{(p)}_{it},
\end{equation}

where $\epsilon^{(p)}_{it}$ is the process error, namely the difference between the interpolated dynamics, $\partial \tilde{y}_i/\partial t$ and the NODE, $dy_i/dt$, given the interpolated state variables $\tilde{y} = \{\tilde{y}_1,\tilde{y}_2, ...,\tilde{y}_I\}$ (Fig. 1, blue boxes). 

\textbf{Bayesian regularisation}

In the context of standard gradient matching, defining the observation model (Eq. 3) and process model (Eq. 6) would be sufficient to fit the NODE system (Eq. 1) to the time series via optimisation (\cite{Jost2000,Ellner2002,Wu2005}).
We could find the parameter vector $\omega_i$ and $\theta_i$ that minimise the sum of squared observation and process errors, $\epsilon_{it}^{(o)}$ and $\epsilon_{it}^{(p)}$ (Eq. 3 and 6).
However, this approach is prone to overfitting, and does not provide estimates of uncertainty around model predictions. 
To account for this, we introduce Bayesian regularisation, which allows us to control for overfitting by constraining parameters with prior distributions (\cite{Cawley2007}), and to root our interpretation of uncertainty in a Bayesian framework.

First, we define a simple Bayesian model to fit the interpolation functions (Eq. 3) to the time series data.
We assume normal distributions for the observation error, $\epsilon^{(o)}_{ij} \sim \mathcal{N}(0,\sigma_i)$, and for the parameters, $\omega_{ij} \sim \mathcal{N}(0,\gamma_{ij})$.
Here, we are only interested in interpolating the time series accurately, irrespective of the value of $\sigma_i$ and $\gamma_{ij}$.
Therefore, we use the approach developed by Cawley and Talbot to average out the value of the parameters $\sigma_i$ and $\gamma_{ij}$ in the full posterior distribution (\cite{Cawley2007}), assuming gamma hyperpriors $p(\xi) \propto \frac{1}{\xi} \exp\left\{- \xi \right\}$ for both parameters.
This yields the following expression for the log marginal posterior density of the parameters,

\vspace{-0.5cm}
\begin{equation}
    \log P(\omega_i ~|~ Y_i) \propto - \frac{J}{2} \log \left(1 + \sum_{j=1}^{J} \left( \epsilon^{(o)}_{ij} \right)^2 \right) - \frac{K}{2} \log \left(1 + \sum_{k=1}^{K} \omega_{ik}^2 \right)
\end{equation}

where $P$ is the marginal posterior density,
$\omega_i = \{\omega_{i1},\omega_{i2},...,\omega_{iK}\}$ is the observation parameter vector controlling the interpolation function,
$Y_i = \{Y_{i1},Y_{i2},...,Y_{iJ}\}$ corresponds to the sequence of observations of state variable $i$ at time step $j$, 
$J$ is the total number of time steps in the time series, 
$\epsilon^{(o)}_{ij}$ is the observation error at time step $j$ between the interpolated and observed  value of variable $i$, 
$K$ is the total number of parameters. 
More details on how to derive this expression can be found in a supplementary file (Supplementary A).

Then, we define a simple Bayesian model to fit the NODEs to the interpolated dynamics, given the interpolated states.
We assume normal distributions for the observation error, $\epsilon^{(p)}_{ij} \sim \mathcal{N}(0,\sigma_i)$, and parameters, $\theta_{ik} \sim \mathcal{N}(0,\delta_{ik})$.
This gives the following expression for the log posterior density of the parameters given the interpolations,

\vspace{-0.5cm}
\begin{equation}
    \log p(\theta_i ~|~ \omega) \propto - \frac{1}{2} \sum_{J=1}^{J} \left( \frac{\epsilon^{(p)}_{ij}}{\sigma_i} \right)^2 - \frac{1}{2} \sum_{k=1}^{K} \left( \frac{\theta_{ik}}{\delta_{ik}} \right)^2
\end{equation}

where $\theta_i = \{\theta_{i1},\theta_{i2},...,\theta_{iK}\}$ are the NODE parameters of the $i^{th}$ variable,
$\omega = \{\omega_1,\omega_2,...,\omega_I\}$ are the interpolation parameters of each state variable, 
$\epsilon^{(p)}_{ij}$ is the process error of variable $i$ at time step $j$ between the interpolated dynamics and NODE prediction, 
$\sigma_i$ is the standard deviation of the likelihood, 
$K$ is the total number of parameters, 
$\delta_{ik}$ is the standard deviation of the prior distribution of parameter $\theta_{ik}$.

This approach allows us to limit overfitting by adjusting the constraint on the parameters, which is controlled by the standard deviation of the parameter prior distributions, $\delta_{ik}$ (\cite{Cawley2007, Bonnaffe2021a}).
We could set small values of $\delta$ to limit the degree of nonlinearity in the response, or to eliminate specific variables from the model by constraining their parameters to be close to zero.
We identify the appropriate degree of constraint $\delta_{i}$ on NODE parameters via cross-validation. 
We train the NODE model on the first half of the interpolated data and predict the remaining half.
We repeat this process for increasing values of $\delta_{i}$, until we find the value that maximises the log likelihood of the test data.

\subsection{Inference and uncertainty quantification}

Finally, we estimate uncertainty in parameter values by anchored ensembling, which produces approximate Bayesian estimates of the posterior distribution of the parameters (\cite{Pearce2018}).
This involves sampling a parameter vector from the prior distributions, $\theta_{i} \sim \mathcal{N}(0,\delta_{i})$, and then optimising the posterior distribution from this starting point, $\theta^*_i = \underset{\theta_i}{argmax}~\log p(\theta_i~|~\omega)$.
By repeatedly taking samples, the sampled distribution $\theta^*$ approaches the posterior distribution and provides estimates and error around the quantities that can be derived from the models.
The expectation and uncertainty around derived quantities can then be obtained by computing the mean and variance of the approximated posterior distributions.
The great strength of this approach is that it is unlikely to get stuck in local maxima hence providing a more robust optimisation of the posterior.

\subsection{Analysing NODEs}

In this study we are mainly interested in two outcomes of NODEs, namely inferring the direction (or effect) and strength (or contribution) of interactions between the state variables (\cite{Bonnaffe2021a}).
We define the direction of the interaction between variable $y_i$ and $y_j$ as the derivative of the dynamics of $y_i$ with respect to $y_j$, and vice versa (\cite{Sugihara2012}), 

\vspace{-0.5cm}
\begin{equation}
    e_{ijt} = \frac{\partial}{\partial y_j} \frac{dy_i}{dt}.
\end{equation}

Knowing the direction, however, is not sufficient to determine the importance of a variable for the dynamics of another. 
Given the same effects, a variable that fluctuates a lot will have a greater impact on the dynamics of a focal variable, compared to a variable that remains quasi-constant.
We hence compute the strength of the interaction by multiplying the dynamics of a variable $y_j$ by its effect on the focal variable $y_i$, also known as the Geber method (\cite{Hairston2005}),

\vspace{-0.5cm}
\begin{equation}
    c_{ijt} = \frac{dy_j}{dt} \frac{\partial}{\partial y_j} \frac{dy_i}{dt}.
\end{equation}

To summarise results across the entire time series we can compute the mean effects $e_{ij}$ by averaging $e_{ijt}$ across all time steps, $e_{ij} = 1/K \sum_k e_{ijk}$, as well as the relative total contribution, $c_{ij}$, of a variable to the dynamics of another by computing the relative sum of square contributions, $c_{ij} = \left( \sum_{ijk} c_{ijk}^{2} \right)^{-1} \sum_t c_{ijt}^2$. 
By computing the direction and strength of interactions between all the variables in the system we can build dynamically informed ecological interaction networks (e.g. fig. 5).
Other metrics can be computed by analysing the NODEs, such as equilibrium states, these are discussed in our previous work (\cite{Bonnaffe2021a}). 

\section{Case studies}

\subsection{Case study 1: artificial tri-trophic prey-predator oscillations}

In this first case study, we aim to demonstrate the accuracy of the NODE fitted by BNGM in inferring nonlinear per-capita growth rates in a system where ground truth is known.
Hence, we simulate a set of time series from a tri-trophic ODE model with known equations and parameters, and we compare the fitted NODEs to the actual ODEs.

\textbf{System}

We consider a tri-trophic ODE system consisting of a prey, an intermediate predator, and a top predator.
The system is built on the real tri-trophic system featuring algae, flagellates, and rotifers, considered in case study 2 (\cite{Hiltunen2013}),

\vspace{-0.5cm}
\begin{equation} \begin{aligned}
    & \frac{dG}{dt} = \left( \alpha \left(1-\frac{G}{\kappa}\right) - \frac{\beta B}{1+\delta G} - \frac{\gamma R}{1+\delta G} \right) G\\
    & \frac{dB}{dt} = \left( \frac{\beta G}{1+\delta G} - \phi R - \mu \right) B \\
	& \frac{dR}{dt} = \left( \frac{\gamma G}{1+\delta G} + \phi B - \nu \right) R,
\end{aligned} \end{equation}

where $G$, $B$, and $R$, correspond to the prey, intermediate, and top predator population densities, respectively,
$\alpha$ is the prey intrinsic growth rate, limited by a carrying capacity $\kappa$, 
$\beta$ and $\gamma$ are the predation rates by the intermediate and top predator,
$\delta$ is the saturation rate of prey predation, which emulates the capacity of the algae to display predator defense at higher algal density (\cite{Hiltunen2013}),
$\phi$ is the predation rate of the intermediate predator by the top predator,
$\mu$ and $\nu$ are the intrinsic mortality of the intermediate and top predator.

We simulate a case of invasion, by introducing the top predator at a low density, with a set of parameters that results in dampening prey-predator oscillations, namely $\alpha = 1$, $\beta = 2.5$, $\gamma = 1.5$, $\kappa = 3$, $\delta = \phi = \mu = \nu = 1$.
We focus on the middle section of the time series, $t \in [20,50]$, as in the initial section the top predator is rare, and in the later section populations have attained a fixed equilibrium point.
The resulting time series are presented in figure 2.

\textbf{NODE model}

In order to nonparametrically learn the per-capita growth rate of each species, and to derive ecological interactions, we define a three-species NODE system,

\vspace{-0.5cm}
\begin{equation} \begin{aligned}
	& \frac{dR}{dt} = r_R(R,G,B,\beta_R) R \\
	& \frac{dG}{dt} = r_G(R,G,B,\beta_G) G \\
	& \frac{dB}{dt} = r_B(R,G,B,\beta_B) B,
\end{aligned} \end{equation}

where the per-capita growth rates $r_R$, $r_G$, and $r_B$ are neural network functions of the density $R$, $G$, $B$ of each species (function $f_p$, Eq. 2).
We choose a combination of linear and exponential activation functions $f_{\sigma, j\leq J/2}(x) = x$, and $f_{\sigma, j>J/2}(x) = \exp(x)$.
This allows us to progressively switch from a simple linear model to a nonlinear model by releasing the constraint on the exponential section of the neural network during cross-validation.
The number of units in the hidden layer $J$ is chosen to be 10, as this is a commonly used number for systems of that size (e.g. \cite{Wu2005,Bonnaffe2021a}). 

\textbf{Time series interpolation}

We interpolate the time series using the neural network described in section 2.3 (Eq. 4).
We set the number of neurons in the network to $J=30$.
We use sinusoid activation functions, $f_\sigma(x) = sin(x)$, so that the weights $\omega^{(1)}_{ij}$, $\omega^{(2)}_{ij}$, and $\omega^{(3)}_{ij}$ control the amplitude, shift, and frequency of the oscillations in the time series, respectively.
Given that the population densities are strictly positive $R,~G,~B \in \mathcal{R^{+}}$, we use an exponential link function, $f_\lambda(x) = \exp (x)$. 
We then approximate the marginal posterior distribution of the interpolation parameters, and thereby of interpolated states and dynamics, by taking 100 samples from the log marginal posterior distribution (Eq. 7) via anchored ensembling. % $\omega^*_{i} = \underset{\omega}{argmax} ~\log P(\omega_i ~|~ Y_i)$, where $Y_i$ denotes the observed time series of either $R$, $G$, and $B$.
In practice, the high number of parameters in the neural network equation may impede the fit of the time series, especially for short time series. 
We found that dividing the number of parameters $K$ (Eq. 7) by the number of neurons in the network $J$ (Eq. 2) yields consistent fitting results.
Interpolated states and dynamics are presented in figure 2.

\textbf{Fitting NODEs to the interpolated time series}

We fit the NODE system to the interpolated time series.
In practice, we fit the NODE to the expectation of the interpolated state and dynamics, $\mathrm{E}(\tilde{y}_i)$ and $\mathrm{E}(d\tilde{y}_i/dt)$, by averaging over all sampled interpolation parameters.
An alternative approach could be to consider the interpolation that maximises the log marginal posterior density, but this may decrease repeatability due to the difficulty of reliably identifying a global maximum.
Averaging across multiple interpolations ensures an overall smoother and robust interpolation. 
In addition, we standardise the response and explanatory variables with respect the their mean and standard deviation (i.e. $Z=(Y-\mu)/\sigma$).
This is to facilitate the training of the NODE by equalizing the scale of the different parameters in the neural network.
Then, we identify the optimal regularisation parameter $\delta$ (Eq. 8) by cross-validation.
To do that, we split the data in half, train NODEs on the first half, and calculate the log likelihood of the test set for increasing values of $\delta$, from $0.05$ (linear) to $0.5$ (highly nonlinear), by increments of $0.05$.
This allows us to identify the maximum degree of nonlinearity, $\delta$, in the per-capita growth rate that ensures generalisability throughout the time series.
Then, we approximate the posterior distribution of the NODE parameters by taking 30 samples from the posterior distribution (Eq. 8).
Finally, we perform model selection by removing variables that do not result in a significant decrease in the log likelihood of the model (assessed by comparing log likelihood confidence intervals).
We ensure moderate temporal autocorrelation and normality by visualising the residuals of the models.
We also ensure results repeatability by running the entire fitting process a second time.

\textbf{Computing ecological interactions}

Finally, we analyse the shape of the per-capita growth rates to recover the interaction between the three species in the system.
In particular, we look at the effect and contribution of each species to the dynamics of the others.
The effect is computed as the sensitivity (i.e. the gradient) of the per-capita growth rate of a given species with respect to the density of the other species (\cite{Sugihara2012,Bonnaffe2021a}).
The contribution is computed following the Geber method (\cite{Hairston2005}), which consists in multiplying the dynamics of a variable by its effects on the other variables.
We further compute the importance of a species in driving the dynamics of another by computing its relative total contribution compared to other species.
More details on how to compute these quantities can be found in section 2.5 and in our previous study (\cite{Bonnaffe2021a}).

\subsection{Case study 2: real tri-trophic prey-predator oscillations}

In this second case study, we want to assess the quality of the NODE analysis when performed on a real time series.
We are further interested in comparing the direction and strength of uncovered ecological interactions across virtually identical replicated time series.

\textbf{System}

We consider a three-species laboratory microcosm consisting of an algal prey (\textit{Chlorella autrophica}), a flagellate intermediate predator (\textit{Oxyrrhis marina}), and a rotifer top predator (\textit{Brachionus plicatilis}).
The algal prey is consumed by the intermediate and top predator, which also consumes the intermediate predator (\cite{Arndt1993}).
The dynamics of this system, here the daily change in the density of each species, were recorded in three replicated time series experiments performed by Hiltunen and colleagues (\cite{Hiltunen2013}). 
We use their time series because they describe a simple yet biologically realistic ecosystem, and because the quality of the replication of their microcosm reduces as much as possible observational and experimental error, and rules out environmental variation (\cite{Hiltunen2013}).
We digitised these time series by extracting by hand the coordinates of every points in the referential of the axis of the graph of the original study, and analysed them.

\textbf{NODE analysis}

We apply the same analysis as performed on the artificial tri-trophic prey-predator oscillations.
This allows us to recover a nonparametric approximation of the growth rate of each species, and then derive the direction and strength of the ecological interactions that underpin their dynamics.
We present detailed results of the analysis of the first time series (Fig. 4), and a summary comparison of the three time series (Fig. 5).
Complementary results, including cross-validation plots, and detailed results for the other two replicates can be found in the supplementary material (Supplementary B-E).

\subsection{Case study 3: real di-trophic prey-predator oscillations}

Finally, we infer ecological interactions by NODE BNGM in the hare-lynx system (\cite{Odum1972}).
This is to provide an example of a longer time series, and to offer a point of comparison with previous and future implementations of NODEs, which commonly use this time series (e.g. \cite{Bonnaffe2021a,Frank2022}).

\textbf{System}

The system is described in details in our previous work (\cite{Bonnaffe2021a}).
The data consist in a 90-year long time series of counts of hare and lynx pelts collected by trappers in the Hudson bay area in Canada (\cite{Odum1972}).
The time series displays characteristic 10-year long prey-predator oscillations.

\textbf{NODE analysis}

We apply the same analysis as previously described, to the exception that the NODE system only features two variables, $H$ and $L$, instead of 3.
Results are presented in figure 6.

\section{Results}

\subsection{Model runtimes}

We present a breakdown of the runtime of fitting NODEs by BNGM for each system in table 1.
We find that it takes on average 5.35 minutes to fit NODEs by BNGM.
This includes taking 390 samples, and thereby performing 390 full optimisations, of the posterior distribution of the interpolation and NODE parameters. 
This amounts to about 5.37 second to sample each variable of the NODE system once.
This is a 335 fold improvement over our previous approach, which took on average 30 minutes (\cite{Bonnaffe2021a}).

%% table
% \newpage
\begin{table}[H]
\begin{center}
\setstretch{1.0}
\caption{
\textbf{Summary of model runtimes.}
We measured the time required to perform 100 interpolations and 30 NODE fits to each variable in the systems.
Replicate A, B, and C correspond to each replicated time series of the aglae, flagellate, and rotifer tri-trophic system (\cite{Hiltunen2013}).
The Hare-Lynx system correspond to the 90 years long time series of hare and lynx pelt counts (\cite{Odum1972}).
The number of time steps (N steps) is given for each time series. 
The total time per fit is obtained by dividing the total time in seconds by the number of fits (i.e. 130).
It takes on average 5.35 minutes for the 130 NODE fits, which amounts to 5.37 seconds per sample taken. 
These results were obtained on a macbook pro M1 MAX 2022, in base R (v4.0.2).
}
\begin{tabular}{rcccccccc}
\hline
& \\
& & & \multicolumn{2}{c}{Interpolation} & \multicolumn{2}{c}{NODE fit} & &  \\
& & & \multicolumn{2}{c}{-------------------------} & \multicolumn{2}{c}{-------------------------} & &  \\
% & \\
System & N var. & N steps & N fits & time (s) & N fits & time (s) & total & total p. fit \\
& \\
\hline
& \\
Replicate A & 3 & 66 &  100 & 239.47 & 30 & 129.41 & 368.88 & 6.71 \\
Replicate B & 3 & 66 &  100 & 233.59 & 30 & 133.13 & 366.72 & 6.77 \\
Replicate C & 3 & 40 &  100 & 136.51 & 30 &  74.01 & 210.52 & 3.83 \\
Hare-lynx   & 2 & 90 &  100 & 303.64 & 30 &  33.56 & 337.20 & 4.16 \\
\end{tabular}
\setstretch{2.0}
\end{center}
\end{table}
\newpage

\subsection{Case study 1: artifical tri-trophic system}

We present the results of fitting NODEs by BNGM to the artificial tri-trophic time series in figure 2 and 3.
We find that both the interpolation of the state variables and dynamics are highly accurate (Fig. 2), given that they closely match the ground truth, known from the equations of the ODE model that we used to generate the time series (Eq. 11).
Similarly, we find that the NODE approximation of the per-capita growth rate of each species also closely matches the ground truth (Fig. 3, a., d., g.). 
We find negative nonlinear effects of the two predators on the growth rate of the algae (Fig. 3, b., blue and purple lines).
This nonlinear pattern is mirrored by the effect of the algae on the growth rate of the predators (Fig. 3, e. and h., red line).
The linear interaction between the two predators is also well-recovered (Fig. 3, e., blue line, and h., purple line).
We find that removing the intra-specific dependence in the growth rate of the predators did not affect the fit of the model (Fig. 3, e., purple line, and h., blue line).
The BNGM approach hence accurately recovers the dynamical characteristics of the artificial system.

%% figure
\newpage
\begin{figure}[H]
\begin{center}
\includegraphics[width=\linewidth,page=2]{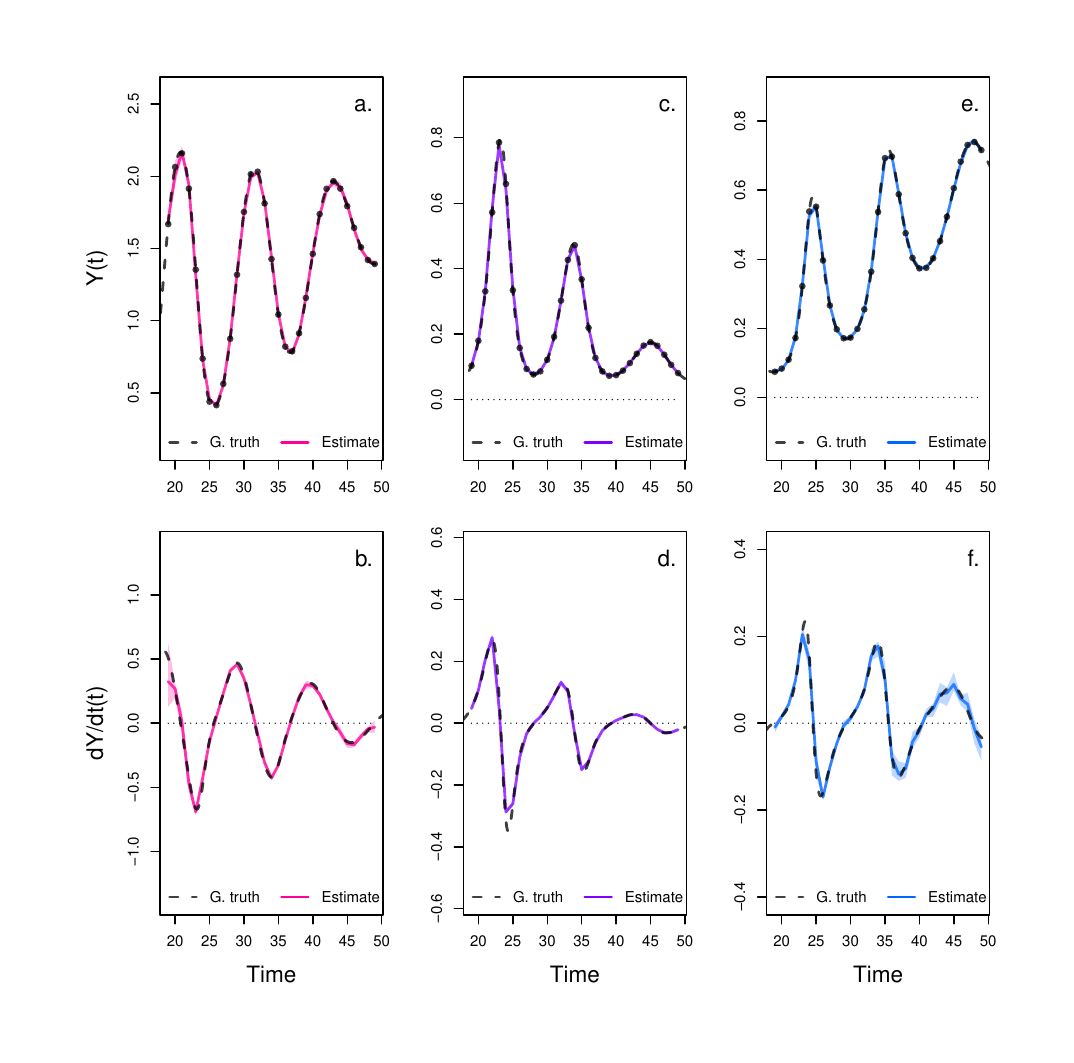}
\caption{
    \textbf{Interpolated density and dynamics of prey, intermediate, and top predators in the artificial system.}
    This figure corresponds to the first step in the overview figure (Fig. 1).
    It shows the accuracy of the interpolated densities of prey (a.), intermediate (c.), and top predators (e.).
    We obtain interpolated densities by fitting observed densities (black dots) with ANNs that take time as input.
    The observed densities were obtained by sampling a tri-trophic prey-predator ODE model at regular time steps.
    We then derive interpolated dynamics (b., d., f.) by computing the temporal derivative of the interpolated densities with respect to time.
    In all graphs, the dashed line represents the ground truth, namely trajectories generated by the ODE model.
    The solid lines correspond to the interpolations. 
    The shaded area shows the 90\% confidence interval, obtained by approximately sampling the marginal posterior distributions. 
}
\end{center}
\end{figure}
\newpage

%% figure
\newpage
\begin{figure}[H]
\begin{center}
\includegraphics[width=\linewidth,page=3]{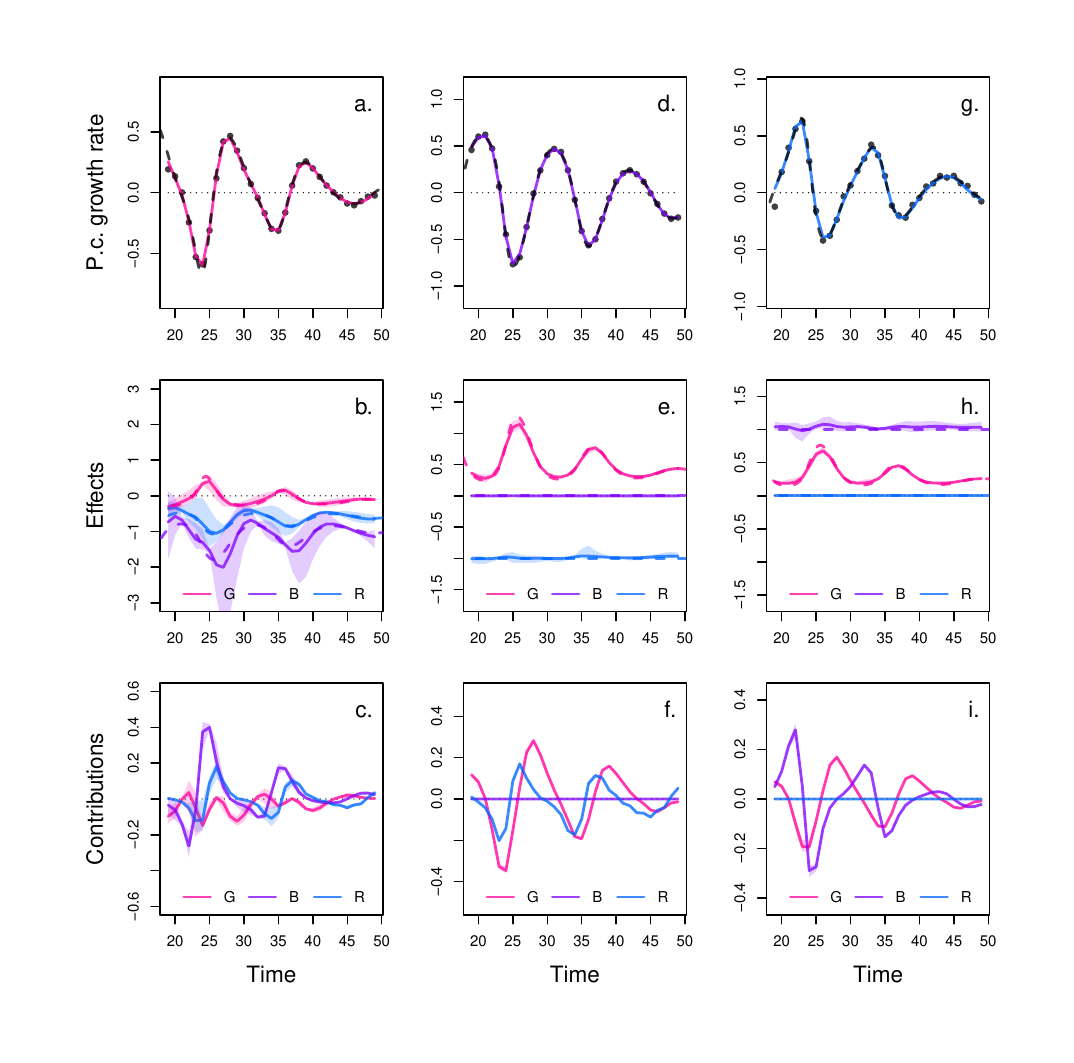}
\caption{
    \textbf{Drivers of dynamics of prey, intermediate, and top predator in the artificial system.}
    This figure corresponds to the second step in the overview figure (Fig. 1).
    It displays the NODE nonparametric approximations of the per-capita growth rate of prey (a., b., c.), intermediate (d., e., f.), and top predators (g., h., i.).
    We obtain the NODE approximations (a., d., g., solid line) by fitting the interpolated per-capita growth rates (black dots) with ANNs that take population densities as input.
    We then estimate the direction of ecological interactions (effects, b., e., h.) by computing the derivative of the NODE approximations with respect to each density.
    Finally, we compute the strength of ecological interactions (contributions, c., f., i.) by multiplying the interpolated dynamics of each population (Fig. 1, b., d., f.) with its effects.
    Dashed lines correspond to ground truth, obtained from the original trajectories of the tri-trophic ODE model. 
    The shaded area shows the 90\% confidence interval, obtained by approximately sampling the posterior distributions. 
}
\end{center}
\end{figure}
\newpage

\subsection{Case study 2: real tri-trophic prey-predator oscillations}

We present the in-depth analysis of the drivers of the dynamics of the algae, flagellate, and rotifer population in replicate A (Fig. 4).
Cross-validation reveals that there is no support for nonlinear effects in the growth rate of the algae and flagellate for replicate A (Fig. 4, a. and b., d. and e.). 
We find negative linear intra-specific density dependence in algal growth (Fig. 4, b., red line), and negative linear inter-specific effects of the two predators (purple and blue line).
We find that the growth rate of the flagellate is virtually solely driven by predation by the rotifer (Fig. 4, e. and f., blue line).
The rotifer population itself is driven by a positive nonlinear effect of both preys (Fig. 4, h., red and purple line).
There is also evidence for positive nonlinear intra-specific density dependence (Fig. 4, h., blue line).
Overall, comparing results across the three replicates reveals that the effect of the rotifer population on the flagellate and algae, and the effect of the algae on the rotifer, are the strongest and most consistent interactions (Fig. 5, table 2).
The interactions of the flagellate with the algae, and its effect on the rotifer population varies substantially across replicates (Fig. 5, table 2). 
Interestingly, intra-specific density dependence in rotifer and algae is also found to be inconsistent across the three replicates.

%% figure
\newpage
\begin{figure}[H]
\begin{center}
\includegraphics[width=\linewidth,page=5]{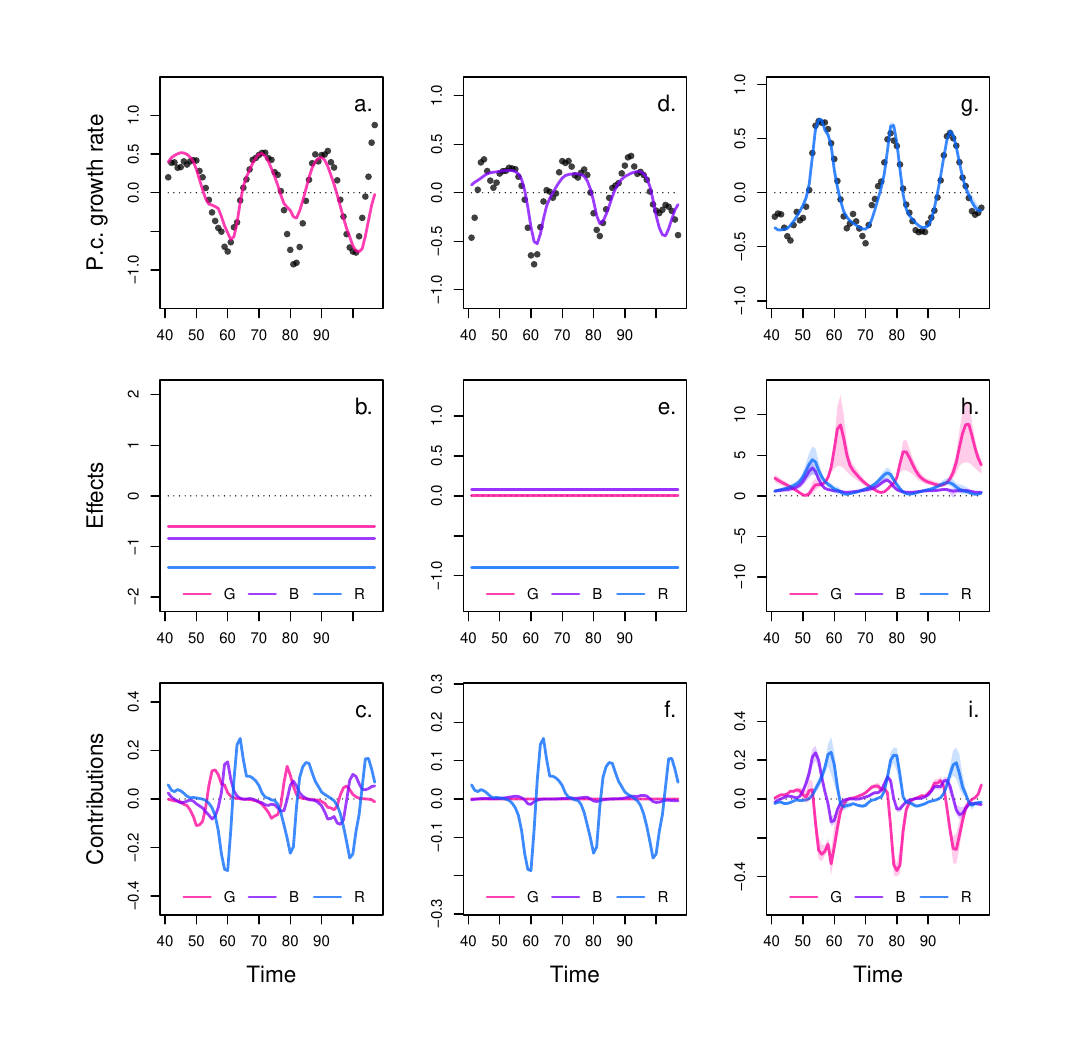}
\caption{
    \textbf{Drivers of dynamics of algae, flagellate, and rotifer in replicate A.}
    This figure displays the NODE nonparametric approximations of the per-capita growth rate of algae (a., b., c.), flagellate (d., e., f.), and rotifer (g., h., i.).
    We obtain the NODE approximations (a., d., g., solid line) by fitting the interpolated per-capita growth rates (black dots) with ANNs that take population densities as input.
    We then estimate the direction of ecological interactions (effects, b., e., h.) by computing the derivative of the NODE approximations with respect to each density.
    Finally, we compute the strength of ecological interactions (contributions, c., f., i.) by multiplying the interpolated dynamics of each population with its effects.
    The shaded area shows the 90\% confidence interval, obtained by approximately sampling the posterior distributions. 
    The replicated time series were obtained by digitising the time series in Hiltunen et al. (2013).
}
\end{center}
\end{figure}
\newpage

%% figure
\newpage
\begin{figure}[H]
\includegraphics[width=1\linewidth,page=6]{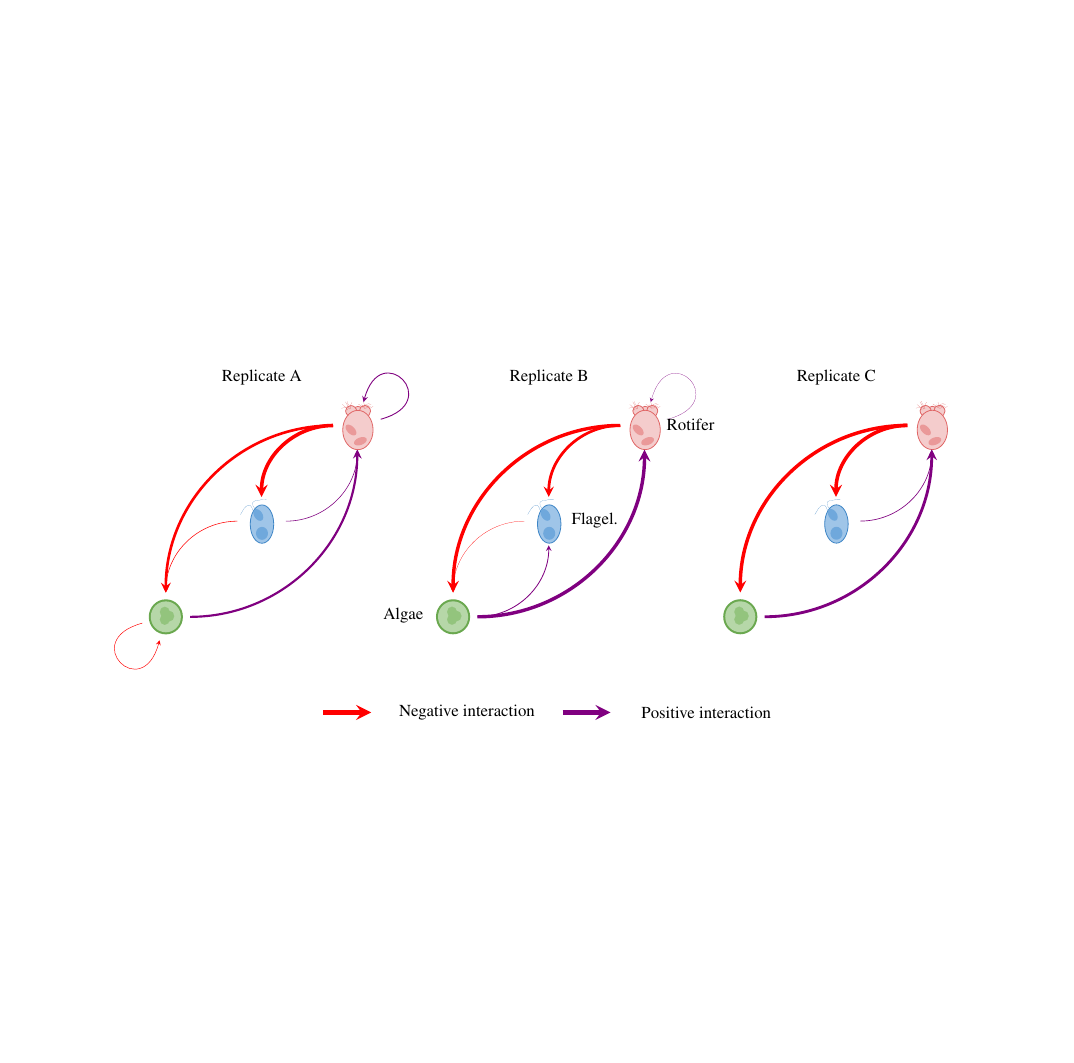}
\vspace{-4cm}
\caption{
    \textbf{Interaction networks inferred from 3 replicated time series of algae, flagellate, and rotifers.}
    This figure shows the direction and strength of ecological interactions inferred from 3 replicated sets of time series of algae, flagellate, and rotifer, using NODEs fitted by BNGM.
    The replicates B and C were analysed in the same way as replicate A (see fig. 4 for details).
    Red and purple arrows correspond to negative or positive mean effects. 
    We estimated mean effects by averaging effects (i.e. derivative of NODE approximated per-capita growth rates with respect to each population density) across the time series.
    The width of the arrows is proportional to the relative strength of the ecological interaction. 
    We compute the relative strength as the \% of total contributions attributable to either algae, flagellate, or rotifer, obtained from summing the square of contributions of each species throughout the time series.
    For instance in replicate A, the relative strength of the effect of rotifer on algae is found by summing the square of the blue line in fig. 4 c., and comparing it to the sum of square of all contributions (Fig. 4 c., red, purple and blue lines).
    We provide the value of the mean effects and relative strengths in table 2.
    The replicated time series were obtained by digitising the time series in Hiltunen et al. (2013).
}
\end{figure}
\newpage

%% table 
\newpage
\begin{table}[H]
\begin{center}
\setstretch{1.0}
\caption{
    \textbf{Comparison of the direction and strentgh of ecological interactions estimated by BNGM across 3 replicated tri-trophic microcosms.}
    Mean effects are obtained by averaging the effect of one species on the growth rate of another throughout the time series.
    The \% of total contributions is obtained by summing the square of contributions of one species density to the growth of the other at each time step throughout the time series, then by computing the proportion of total change that it accounts for.
    The variables $G$, $B$, and $R$ correspond to the population density of algae, flagellate, and rotifer respectively.
    $r^2$ corresponds to the r squared of the NODE nonparametric approximation of the pre-capita growth rate of the three species.
}
\begin{tabular}{rrcccc}
	\hline
	\\
	& & & G & B & R \\
	& \\
	\hline
	& \\
	% & \textbf{replicate A} \\
	& \textbf{Replicate A} & $r^2$ & 0.3 & 0.47 & 0.94 \\
	& \\
	& \textbf{Mean effects} 
	&   on G &  -0.61 & -0.85 & -1.41 \\
	& & on B &   0.00 &  0.08 & -0.90 \\
    & & on R &   2.84 &  0.93 &  1.23 \\
	& \\
	& \textbf{\% of total contributions} 
	&   to G &   0.13 &  0.15 &  0.73 \\ 
    & & to G &   0.00 &  0.00 &  1.00 \\
    & & to R &   0.60 &  0.16 &  0.25 \\
	& \\
	\hline
	& \\
	% & \textbf{replicate B} \\
    & \textbf{Replicate B} & $r^2$  &  0.65 & 0.85 & 0.47 \\
	& \\
	& \textbf{Mean effects} 
    &   on G &  0.00 & -0.56 & -1.13 \\
	& & on B &  0.34 &  0.00 & -0.58 \\
	& & on R &  0.87 &  0.00 &  0.19 \\
	& \\
	& \textbf{\% of total contributions} 
    &   to G &  0.00 &  0.06 &  0.94 \\ 
    & & to B &  0.23 &  0.00 &  0.77 \\
    & & to R &  0.95 &  0.00 &  0.05 \\
	& \\
	\hline
	& \\
	% & \textbf{replicate C} \\
    & \textbf{Replicate C} & $r^2$  &  0.93 & 0.29 & 0.87 \\
	& \\
	& \textbf{Mean effects} 
    &   on G & -0.14 &  0.13 & -2.31 \\
    & & on B & -0.05 & -0.09 & -0.72 \\
    & & on R &  2.46 &  0.49 & -0.09 \\
	& \\
	& \textbf{\% of total contributions} 
    &   to G &  0.02 &  0.02 &  0.96 \\
    & & to B &  0.00 &  0.01 &  0.99 \\
    & & to R &  0.79 &  0.18 &  0.03 \\
\end{tabular}
\setstretch{2.0}
\end{center}
\end{table}
\newpage

\subsection{Case study 3: real di-trophic prey-predator oscillations}

Finally, we present the analysis of the drivers of the hare-lynx population dynamics in figure 6.
Cross-validation provides support for nonlinear effects in the per-capita growth rate of the hare and lynx.
We find that the hare population growth rate is mostly determined by a nonlinear negative effect of the lynx population (Fig. 6, b. and c. blue line), and by weak nonlinear positive density dependence (red line). 
The lynx growth rate is determined by a positive nonlinear effect of the hare (Fig. 6, e. and f., red line), and to a lesser extent by negative nonlinear intra-specific density dependence (blue line).

%% figure
\newpage
\begin{figure}[H]
\begin{center}
\includegraphics[width=\linewidth,page=4]{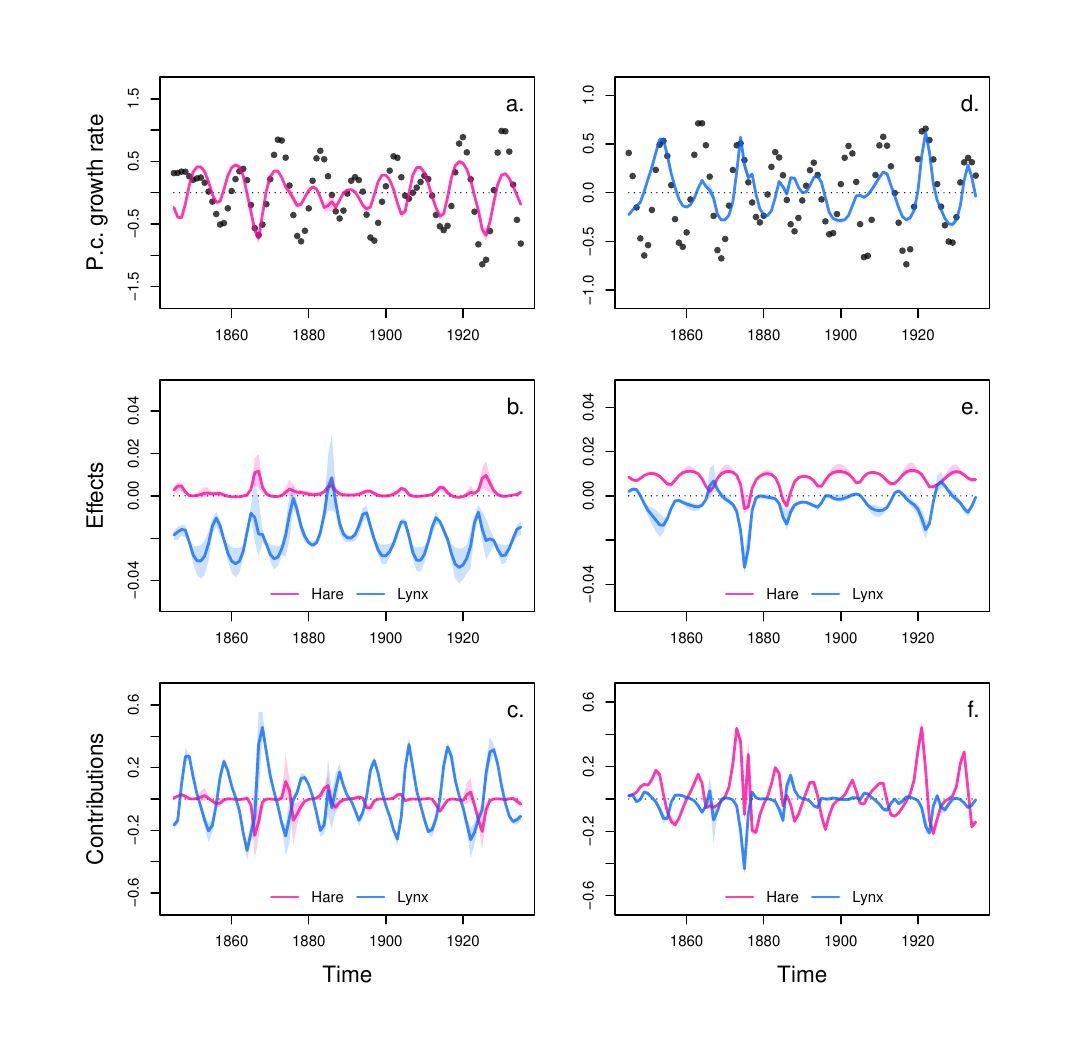}
\caption{
    \textbf{Drivers of dynamics of hare and lynx in the Odum and Barrett pelt count time series.}
    This figure displays the NODE nonparametric approximations of the per-capita growth rate of hare (a., b., c.), and lynx (d., e., f.).
    We obtain the NODE approximations (a., d., solid line) by fitting the interpolated per-capita growth rates (black dots) with ANNs that take population densities as input.
    We then estimate the direction of ecological interactions (effects, b., e.) by computing the derivative of the NODE approximations with respect to each density.
    Finally, we compute the strength of ecological interactions (contributions, c., f.) by multiplying the interpolated dynamics of each population with its effects.
    The shaded area shows the 90\% confidence interval, obtained by approximately sampling the posterior distributions. 
}
\end{center}
\end{figure}
\newpage

\section{Discussion}

Characterising ecological interactions from time series data is challenging.
This is due to the fact that interactions can be highly context-dependent processes (\cite{Song2021}), making it difficult to identify parametric models that encapsulate their complexity (\cite{Wood2001}).
Interactions estimated with parametric models are contingent on the parameterisation arbitrarily chosen by the observer, and hence risk being biased (\cite{Wood2001,Adamson2013}).
We provide a novel method for estimating ecological interactions nonparametrically, by using neural ordinary differential equations (NODEs) fitted with Bayesian neural gradient matching (BNGM). 
First, we remove the cost of fitting NODEs by introducing BNGM, which allows for NODE fitting in only a few seconds. 
The method involves interpolating time series and dynamics with neural networks, and then fitting NODEs to interpolated dynamics with Bayesian regularisation.
We further demonstrate that this approach is accurate, as NODEs approximate with minimal error the ecological interactions in artificial time series, where real interactions are known. 
Finally, we estimate the strength, direction, importance, and nonlinearity of ecological interactions in 4 time series from natural and experimental systems, showing variation in ecological interactions within and across the time series.

\textbf{Bayesian neural gradient matching}

The Bayesian neural gradient matching (BNGM) approach that we propose here extends standard gradient matching, by using artificial neural networks (ANNs) as interpolating functions, and Bayesian regularisation to control the nonlinearity of the processes (\cite{Cawley2007}).
This allows us to accurately fit NODEs within seconds, making it potentially the most efficient current fitting technique available (see also \cite{Treven2021}). 
The use of ANNs as interpolating functions sets it apart from the initial approach of Ellner et al., who use splines to interpolate the time series before approximating the ODEs (\cite{Ellner2002}).
ANNs are more general and flexible than splines, as well as being easier to manipulate given that they are defined continuously on the state space, which is especially useful when handling multiple interactions between variables.
Our approach is related to that of Wu et al., who use ANNs to approximate both the states and ODEs of prey-predator systems (\cite{Wu2005}), as well as that of Treven and colleagues, who developed the Gaussian process equivalent (\cite{Treven2021}).
In both approaches, they train the interpolation functions at the same time as the NODEs, in order to constrain the interpolation of trajectories such that they can be achieved by the NODE system, which thereby introduces dynamical coupling between state variables.
One of the risks of dynamical coupling approaches is that misestimating one of the state variables of the model biases the estimation of the states and dynamics of other variables.
To avoid this, we fit each interpolation and NODE independently to each time series.
In addition, this makes it possible to parallelise the code, resulting in potentially even faster computation.
Our approach removes the main limitation of using NODEs, allowing for quick and extensive model comparison, cross-validation, and uncertainty quantification around estimates.

\textbf{Accuracy of NODEs in estimating ecological interactions}

Our approach relies on approximating population dynamics with NODEs and then computing their sensitivity to a change in the density of the different populations in the system (\cite{Bonnaffe2021a}). 
We demonstrate that NODEs accurately recover the dynamics, strength, direction, and nonlinearity of ecological interactions in artificial tri-trophic prey-predator time series, where truth is known.
In particular, we find that the interactions between the algae and the two predators are nonlinear, and thereby oscillate throughout the time series, which is consistent with the model, that features a resistance to predation at high algal density.
We also recover the linear interactions between the two predators, which shows that the NODEs are sensitive enough to discriminate between linear and nonlinear interactions within and across time series.
To our knowledge, this is the first assessment of the accuracy of NODEs in recovering interactions between variables from time series data, as most of the work focuses on assessing the accuracy of the fitting and forecasting of time series (e.g. \cite{Mai2016,Chen2018,Treven2021,Frank2022}).

\textbf{Ecological interactions in real prey-predator systems}

We further tested NODEs in a real setting, by inferring ecological interactions across three replicated time series of an experimental tri-trophic system of algae, flagellate, and rotifer populations (\cite{Hiltunen2013}). 
Our approach reveals that only stronger interactions, namely the negative effects of the rotifer top predator on the other species, and the positive effect of algae on the rotifer, are conserved across the three replicated time series.
We also find evidence for nonlinearity in the dynamics of the rotifer, as the positive effect of the algae on rotifer growth oscillates throughout the time series. 
This is consistent with the biology of the system, as the algae tends to form anti-predation clumps at higher density, which would dampen the positive effect of algal density on rotifer growth at high algal density (\cite{Yoshida2003,Hiltunen2013}).
We find it interesting that the weaker interactions with the flagellate predator are not consistent across time series, given the controlled laboratory conditions.
This system is known to evolve rapidly, it is hence possible that fast evolution of the different populations from the onset of the time series may have driven the system onto different attractors (\cite{Yoshida2003,Yoshida2007,Hiltunen2013}).
Additionally, stochasticity in population dynamics may have a similar effect (\cite{Dallas2021}).
Disentangling these two sources of variation would require refining the modelling framework, for instance by explicitly including evolution in the model (e.g. with the Price equation, \cite{Ellner2011}), and by using neural stochastic differential equations (i.e. NSDEs, \cite{Rackauckas2019}) fitted with a particle filter. 
While these would constitute interesting developments, our method is still a useful first step, identifying differences between the time series, and demonstrating a reasonable amount of deterministic consistency in the dynamics, judging by the cross-validation and fits.

We also analysed the hare-lynx time series (\cite{Odum1972}), as it is a common benchmark in the field of time series analysis, and provides a comparison point with our previous implementation of NODEs (\cite{Bonnaffe2021a}).
As in our previous study, we found a predatory inter-specific interaction between lynx and hare, and negative intra-specific density dependence in the lynx.
Evidence for positive density dependence in the hare was more limited than previously found.
We also found stronger evidence for nonlinearity, as intra- and inter-specific effects oscillated throughout the time series, as a result of density dependence.
This difference with our previous study is due to the fact that our previous implementation of NODEs was based on simulating the full NODE system, and hence imposed dynamical coupling between the variables.
This dynamical coupling comes at a cost, if one variable is not explained well by the model, it will bias the interactions and dynamics of other variables.
Here, the time series of lynx and hare are analysed independently, each state variable is interpolated as closely as desired, its effects on the dynamics of other variables are hence even more robust to model misspecification than before.

Overall, our approach provides a novel and powerful way of estimating interactions nonparametrically from time series data.
The benefit of using NODEs is that they make no assumptions about the nature of the ecological interactions that drive the dynamics of the species (\cite{Chen2018,Bonnaffe2021a}). 
Hence, we have a better chance at estimating the actual value of the interactions, knowing that it is not subjected to potential incorrect model specifications (\cite{Jost2000,Ellner2002,Wu2005,Kendall2005,Adamson2013}).
This approach is similar to Sugihara's maps (S-maps, \cite{Sugihara2012}), which estimate interactions in time series by approximating the Jacobian matrix nonparametrically via locally linear approximations of the state space (\cite{Deyle2015}).
However, because S-maps are locally linear, they do not assume the existence of a latent trajectory generated by an overarching model.
This creates two caveats, the first being that they are more sensitive to noise in the time series (\cite{Cenci2018}), the second being that they have no theoretical grounding given that they are at heart linear functions defined piecewise on the state space.
NODEs remain in essence deterministic ODE models, assuming an overarching model driving the populations through the entire state-space, which can hence incorporate parametric assumptions regarding the driving processes (\cite{Bonnaffe2021a}).
For instance, we model the per capita growth rate of populations explicitly in NODEs, while S-maps approximate the population-level growth. 
Overall, this makes NODEs more suitable than S-maps for fitting noisy data or exploring theory by testing specific assumptions.

\textbf{Limits and prospects}

One of the main difficulty in quantifying ecological interactions is to identify potential context dependences on other state variables (\cite{Song2021}), for example, whether predation rates are affected by temperature.
Our approach allows for the quantification of context dependence, which shows as nonlinear fluctuations of interactions throughout the time series.
In the present work, we only report nonlinearity as evidence for context dependence in the interactions, but we do not attempt to understand what it is attributable to.
For instance, we identify nonlinear density dependence in the effect of the algae on the rotifer, but we do not know whether this is due to a change in the effect with algae density or rotifer density, or both. 
In order to disentangle these higher order effects we could compute the Hessian of the system, namely the second order derivative of the dynamics with respect to the different state variables. 
Though this procedure is simple mathematically, it would result in 27 second order effects to analyse for the simple 3 species system considered here.
This type of analysis would get rapidly out of hand for larger systems.
Further work should hence consider how to handle these higher order effects, as a way to unveil context dependence in ecological interactions. 

One further issue is that some interactions may depend on variables that are not observed.
For instance, some population dynamics are strongly determined by their demographic state (\cite{Lande2002,Coulson2004}), which would call for time series of the relevant demographic stages.
In the system considered here, the dynamics of algae in the rotifer system are most likely coupled with that of nitrogen, for which no time series was available (\cite{Hiltunen2013}).
Our method only accounts for observed variables, so that time series for all important variables are required, though unaccounted variables are captured to some extent by nonlinear fluctuations in interactions.
One interesting prospect would hence be to incorporate unobserved/latent state variables into the NODE system (\cite{Dupont2019,Zhang2019,Frank2022}).
Careful thought has to be given here as whether to use an ODE or NODE for the latent states given that they are not constrained by observations. 

We consider NODEs, which are only defined along the time dimension.
The framework could easily be extended to any other dimension by considering partial differential equations instead (\cite{Rackauckas2019}).
For instance, in a spatial ecology context we could model the dynamics of populations along two additional spatial dimensions.
In an evolutionary context, we could model the dynamics of populations in phenotype space, by adding phenotypic traits as an additional dimension.
The BNGM method could be instrumental in fitting these models, which are notoriously expensive to stimulate.

\textbf{Conclusion}

We provide a method, BNGM, which allows for NODE fitting in a matter of seconds.
This is a crucial step for efficient model selection and uncertainty quantification in NODEs.
We also demonstrate that NODEs allow for accurate estimation of the direction, strength, and nonlinearity of ecological interactions, in a system where truth is known.
Finally, we estimate ecological interactions in real prey-predator systems, showing that system dynamics are driven by a mixture of linear and nonlinear interactions, of which only strong ones seem to be generalisable across time series.
Our study allows for efficient NODE fitting, and confirms the power of NODEs in identifying dynamical coupling between populations. 

\textbf{Acknowledgments}

We thank warmly the Ecological and Evolutionary Dynamics Lab and Sheldon Lab Group at the department of Zoology for their feedback and support.
We thank Ben Sheldon for insightful suggestions on early versions of the work.
The work was supported by the Oxford-Oxitec scholarship and the NERC DTP.

\textbf{Data accessibility}

All data and code is available at https://github.com/WillemBonnaffe/NODEBNGM.

\textbf{Statement of authorship}

Willem Bonnaff\'e designed the method, performed the analysis, wrote the manuscript; 
Tim Coulson led investigations, provided input for the manuscript, commented on the manuscript.

\setstretch{1.25}
\printbibliography 

% %% figures
% \newpage
% \pagenumbering{gobble}

\newpage
\section{Supplementary}
\appendix
\beginsupplement
\setstretch{1.5}

\section{Bayesian regularisation}

The fitting of the models is performed in a Bayesian framework, considering normal error structure for the residuals, and normal prior density distributions on the parameters

\vspace{-0.5cm}
\begin{equation}
	p(\theta | \mathcal{D}) \propto  p(\mathcal{D} | \theta) p(\theta)
\end{equation}

where $\theta$ is the parameter vector of the model, and $\mathcal{D}$ the evidence, namely the data that the model is fitted to.
% In the case of the interpolation, the evidence is the population densities, either $R(t)$, $G(t)$, or $B(t)$, and the parameters are the weights $\Omega$ in the sinusoid SLPs.
% In the case of fitting the NODE model to the interpolated data, the evidence is the interpolated per-capita growth rate of each population, either $\tilde{r}_R$, $\tilde{r}_G$, or $\tilde{r}_B$, and the parameters are the weights $\beta_R$, $\beta_G$, and $\beta_B$ in the nonparametric per-capita growth rates $r_R$, $r_G$, and $r_B$.
Assuming a normal likelihood for the residuals given the evidence we get

\vspace{-0.5cm}
\begin{equation}
	p( \mathcal{D} | \theta) = \prod_{i=1}^{I} \frac{1}{\sqrt{2\pi\sigma^2}}  \exp \left\{ -\frac{e_i(\mathcal{D},\theta)^2}{2\sigma^2}  \right\}
\end{equation}

where $e_i(\mathcal{D},\theta)$ are the residuals of the model given the parameters, and the evidence. 
In the case of the interpolation, the residuals correspond to the observation error $\epsilon^{(o)}$ (equation 3).
In the case of the NODE approximation, they correspond to the process error $\epsilon^{(p)}$ (equation 7).
% The dispersion term $\sigma$ in the likelihood is measured by the parameters $\sigma_1$ in the case of the interpolation, and $\sigma_2$ in the case of the NODE fitting.
$I$ is the number of data points, either observations in the case of the interpolation, or interpolated points in the case of the NODE fitting.

The prior probability density functions for the parameters are given by

\vspace{-0.5cm}
\begin{equation}
	p(\theta) = \prod_{j=1}^{J} \frac{1}{\sqrt{2\pi\delta^2}}  \exp \left\{ -\frac{\theta_j^2}{2\delta_j^2}  \right\}
\end{equation}

where $J$ is the number of parameters in the models.
The parameter $\delta_j$ controls the dispersion of the priors, and thereby the complexity/level of constraint of the model.

Bayesian regularisation simply amounts to constraining the values of the parameters in the model to be close to a desired value. 
Usually, parameters are constrained by choosing normal priors centered about 0.
In this case, the standard deviation of the normal priors governs the range of values that the parameters can take, and hence constrains more or less strongly the behaviour of the model (\cite{Cawley2007}).
There is no standard approach for choosing $\delta$.
Low values of dispersion may increase constraint on parameters too drastically, which would lead to underfitting, and result in a reduction of the variance of parameter estimates and bias mean estimates towards 0.
In contrast, too high values of dispersion may lead to overfitting, by allowing for more complex shapes.
To account for this, we optimise models on the second-level of inference.
This means that we are finding the optimal value of $\delta$, in addition to optimising the model parameters. 
% Performing inference on the second level means that we are trying to find the appropriate value of the dispersion of the priors, in other words, the appropriate level of constraint on the model. 

In practice, choosing the level of constraint is difficult, Cawley and Talbot hence developed a criterion to perform model selection on the second level of inference.
They proposed to optimise the marginal posterior distribution by averaging out the dispersion of the priors.
With an appropriate choice of prior, the dispersion can be integrated out, leaving us with a formula for the posterior that only depends on the parameters of the model,

\begin{equation}
	\log P(\theta | \mathcal{D}) \propto - \frac{I}{2} \log \left(\sum_{i=1}^{I} e_i(\mathcal{D},\theta)^2\right) - \frac{J}{2} \log \left(\sum_{j=1}^{J} \theta_{j}^2 \right)
\end{equation}

where $P(\theta|\mathcal{D})$ denotes the marginal posterior density, $\mathcal{D}$ denotes the evidence, $I$ and $J$ denote the number of data points and parameters, respectively, $e_i$ denote the residuals, and $\theta$ denote the parameters of the model.
The construction is elegant because it is not sensitive to the choice of prior hyperparameters, and simple as it amounts to optimising the log of the sum of squares, rather than the sum of squares (in the case of normal ordinary least squares).

The issue is that the marginal posterior density is not finite when the parameters are 0, which can lead to underfitting.
In this paper we use a modified criterion, which corrects for that problem,

\begin{equation}
	\log P(\theta | \mathcal{D}) \propto - \frac{I}{2} \log \left(1 + \sum_{i=1}^{I} e_i(\mathcal{D},\theta)^2\right) - \frac{J}{2} \log \left(1 + \sum_{j=1}^{J} \theta_{j}^2 \right)
\end{equation}

where the marginal posterior density depends only on the residuals of the model when the parameters are equal to 0, and otherwise depends on both the parameters and the residudals. 
This construction can be obtained simply by assuming a gamma prior for the parameters $p(\xi) \propto \frac{1}{\xi} \exp\left\{- \xi \right\}$, where $\xi$ is the regularisation parameter, instead of the improper Jeffreys' prior that Cawley and Talbot used in their original study, namely $p(\xi) \propto \frac{1}{\xi}$. 
The details of the integration of the posterior distribution over $\xi$ can be found in Cawley and Talbot's orginal paper.

%%
% \newpage
\section{Complementary results case study 2 replicate A}

%% figure
% \newpage
\begin{figure}[H]
\includegraphics[width=1\linewidth,page=7]{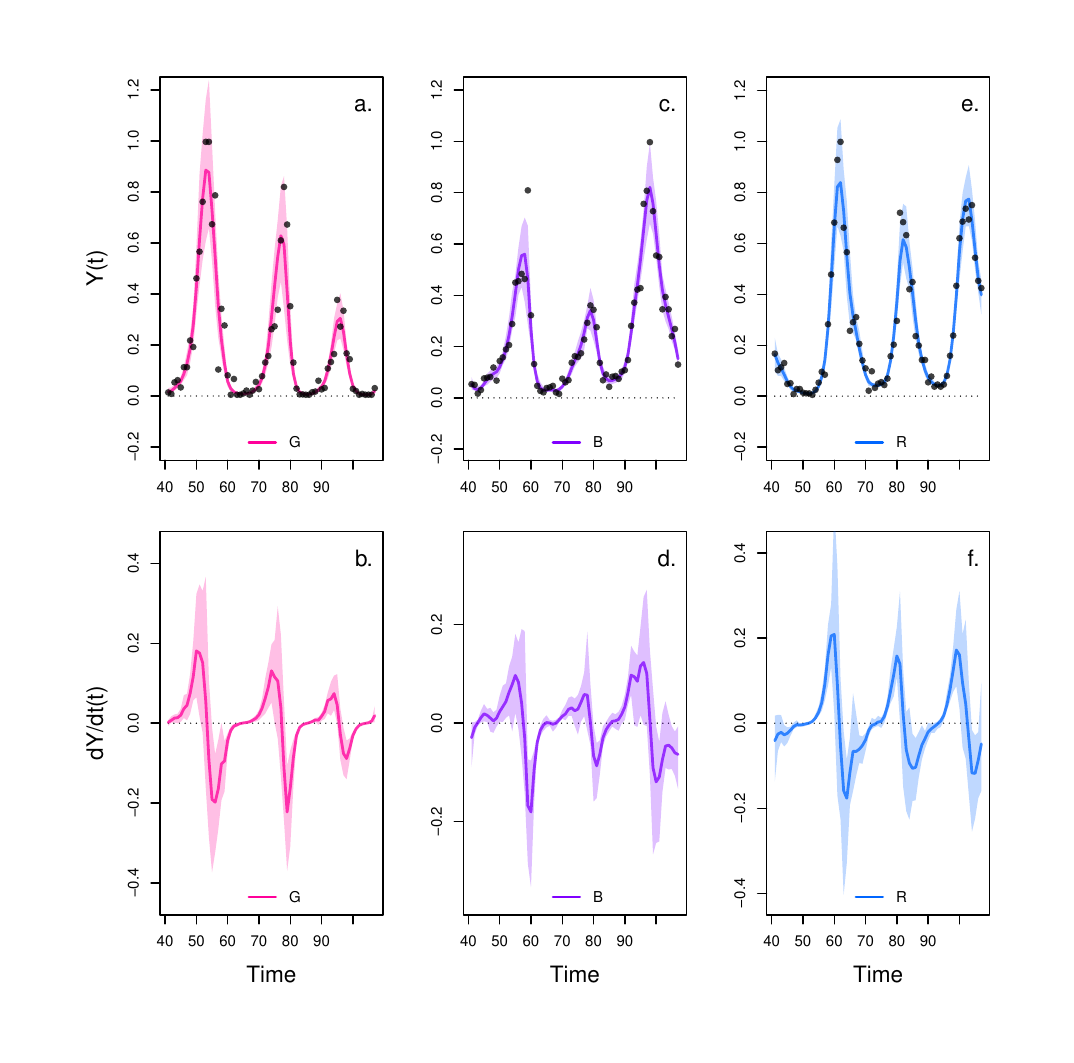}
\caption{
    \textbf{Interpolation of state and dynamics of algae, flagellate, and rotifer density in replicate A.}
    Graph a., c., and e. display the neural interpolation of the population density of algae (G), flagellate (B), and rotifer (R), respectively (obatined with Eq. 7). 
    Graph b., d., and f. show the corresponding interpolated dynamics, obtained by differentiating the interpolation of the states with respect to time (Eq. 5).
    The shaded areas represent the 90\% confidence interval on estimates, obtained by anchored ensembling of the log marginal posterior distribution (Eq. 7) (\cite{Pearce2018}).
    Time series are obtained from digitising the time series in \cite{Hiltunen2013}.
}
\end{figure}
\newpage

%% figure
\newpage
\begin{figure}[H]
\includegraphics[width=1\linewidth,page=8]{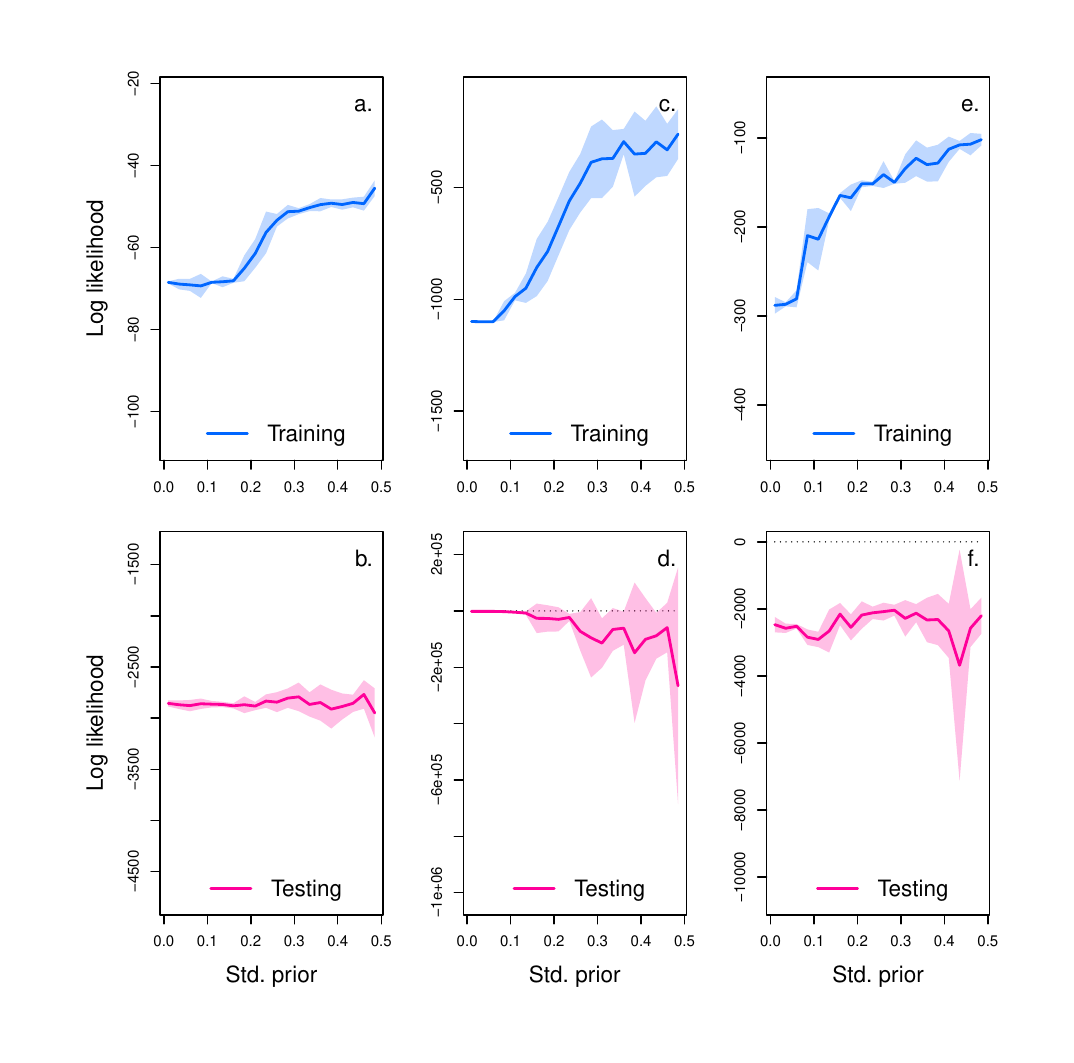}
\caption{
    \textbf{Cross-validation plot of the NODE analysis of replicate A.}
    The x-axis of the graphs correspond to the standard deviation of the prior distribution of the NODE parameters, which constrains the nonlinearity of the nonparametric approximation of the NODEs.
    Small values of standard deviation correspond to a linear model, while higher values (towards 0.5) correspond to a highly nonlinear model.
    Time series of algae, flagellate, and rotifer are split in half to create a train set and a test set. 
    The model is fitted to the train set for increasing value of standard deviation (from 0.05 to 0.5 by 0.05 increments), and evaluted on the test set.
    Graph a., c., and e. show the log likelihood of the NODE system fitted by BNGM to the train set of algae, flagellate, and rotifer, respectively.
    Graph b., d., and f. show the log likelihood of the fitted NODE, evaluated on the corresponding test set.
    The shaded areas represent the 90\% confidence interval on estimates, obtained by anchored ensembling of the log marginal posterior distribution (Eq. 7) (\cite{Pearce2018}).
}
\end{figure}
\newpage

%% figure
\newpage
\begin{figure}[H]
\includegraphics[width=1\linewidth,page=9]{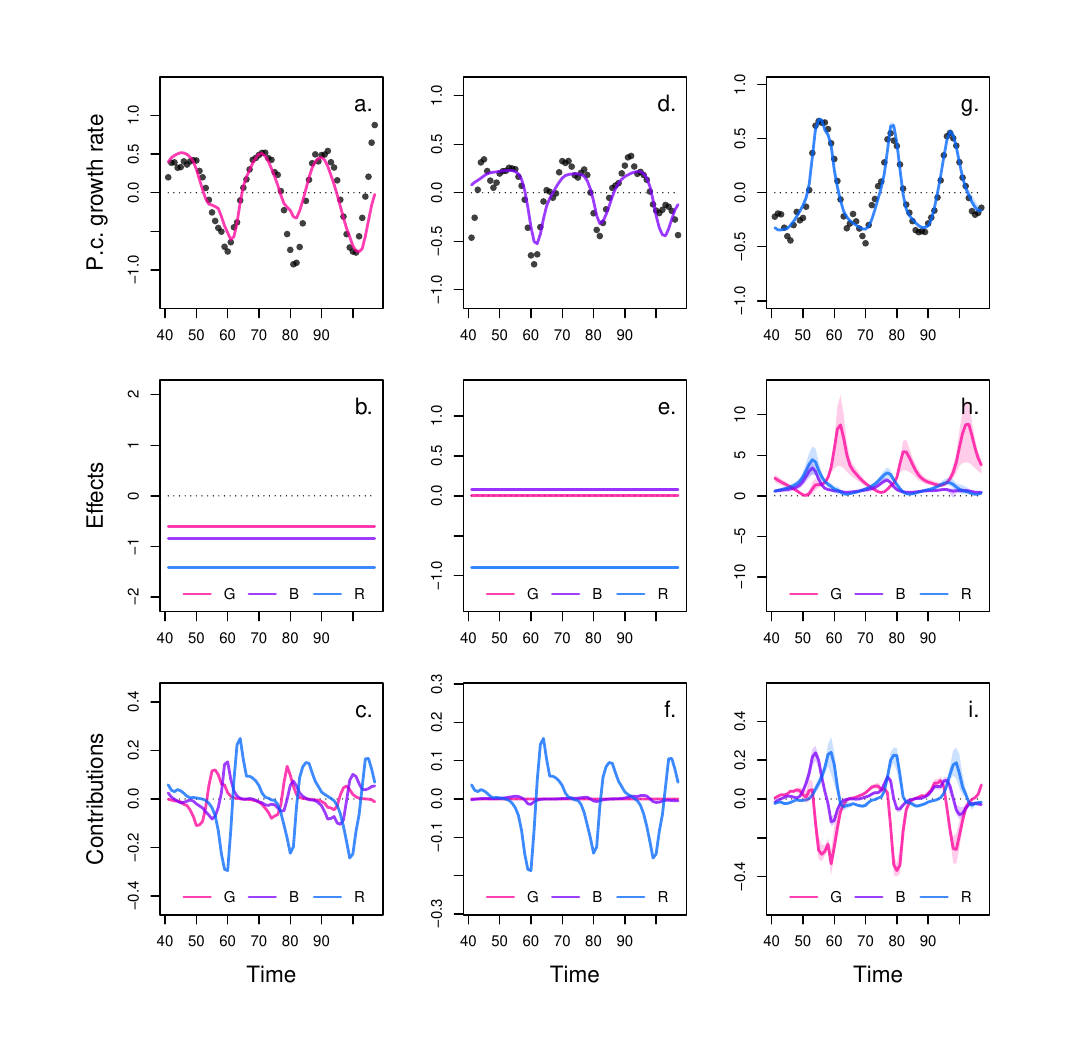}
\caption{
    \textbf{Drivers of dynamics of algae, flagellate, and rotifer in replicate A.}
    This figure displays the NODE nonparametric approximations of the per-capita growth rate of algae (a., b., c.), flagellate (d., e., f.), and rotifer (g., h., i.).
    We obtain the NODE approximations (a., d., g., solid line) by fitting the interpolated per-capita growth rates (black dots) with ANNs that take population densities as input.
    We then estimate the direction of ecological interactions (effects, b., e., h.) by computing the derivative of the NODE approximations with respect to each density.
    Finally, we compute the strength of ecological interactions (contributions, c., f., i.) by multiplying the interpolated dynamics of each population with its effects.
    The shaded area shows the 90\% confidence interval, obtained by approximately sampling the posterior distributions. 
    The replicated time series were obtained by digitising the time series in Hiltunen et al. (2013).
}
\end{figure}
\newpage

%%
% \newpage
\section{Complementary results case study 2 replicate B}

%% figure
% \newpage
\begin{figure}[H]
\includegraphics[width=1\linewidth,page=10]{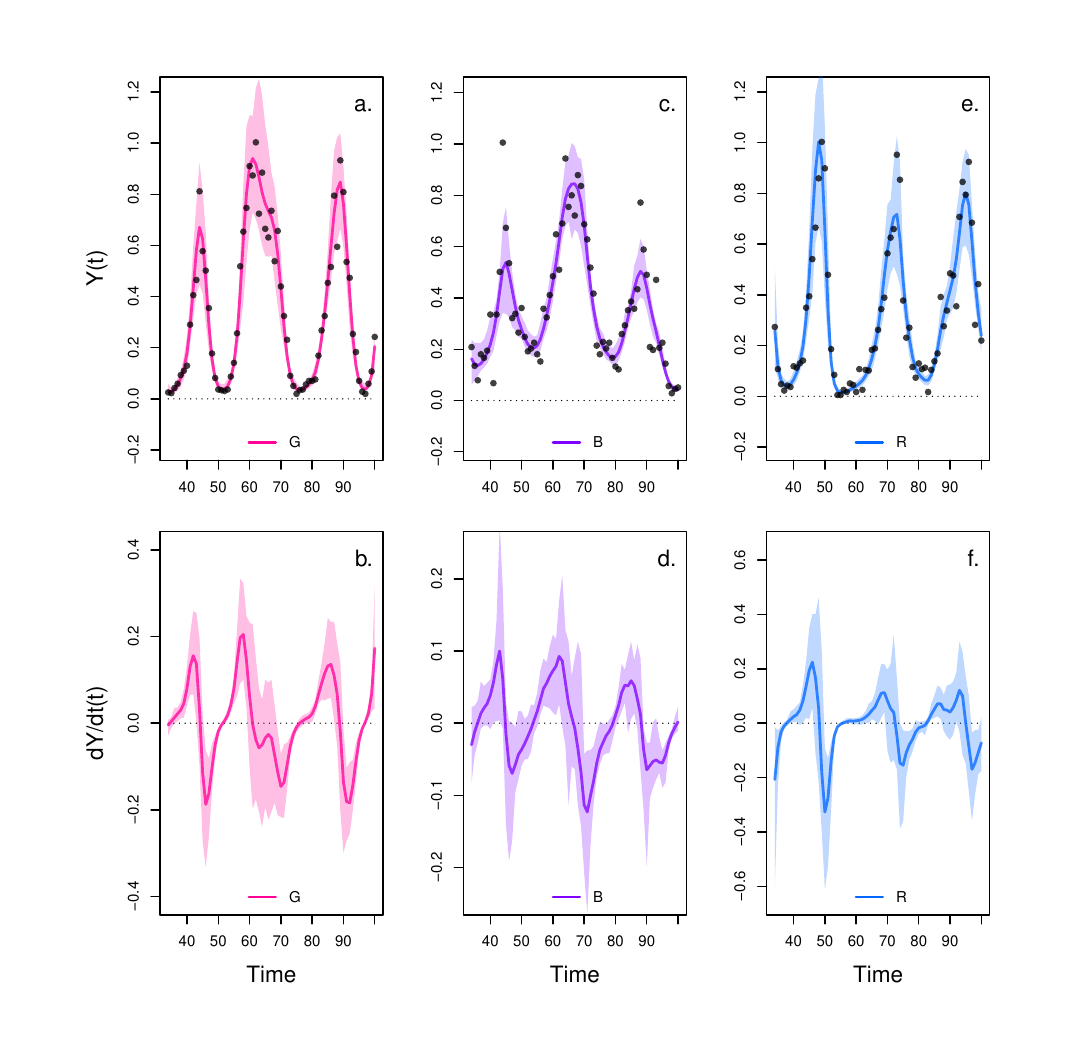}
\caption{
    \textbf{Interpolation of state and dynamics of algae, flagellate, and rotifer density in replicate B.}
    Graph a., c., and e. display the neural interpolation of the population density of algae (G), flagellate (B), and rotifer (R), respectively (obatined with Eq. 7). 
    Graph b., d., and f. show the corresponding interpolated dynamics, obtained by differentiating the interpolation of the states with respect to time (Eq. 5).
    The shaded areas represent the 90\% confidence interval on estimates, obtained by anchored ensembling of the log marginal posterior distribution (Eq. 7) (\cite{Pearce2018}).
    Time series are obtained from digitising the time series in \cite{Hiltunen2013}.
}
\end{figure}
\newpage

%% figure
\newpage
\begin{figure}[H]
\includegraphics[width=1\linewidth,page=11]{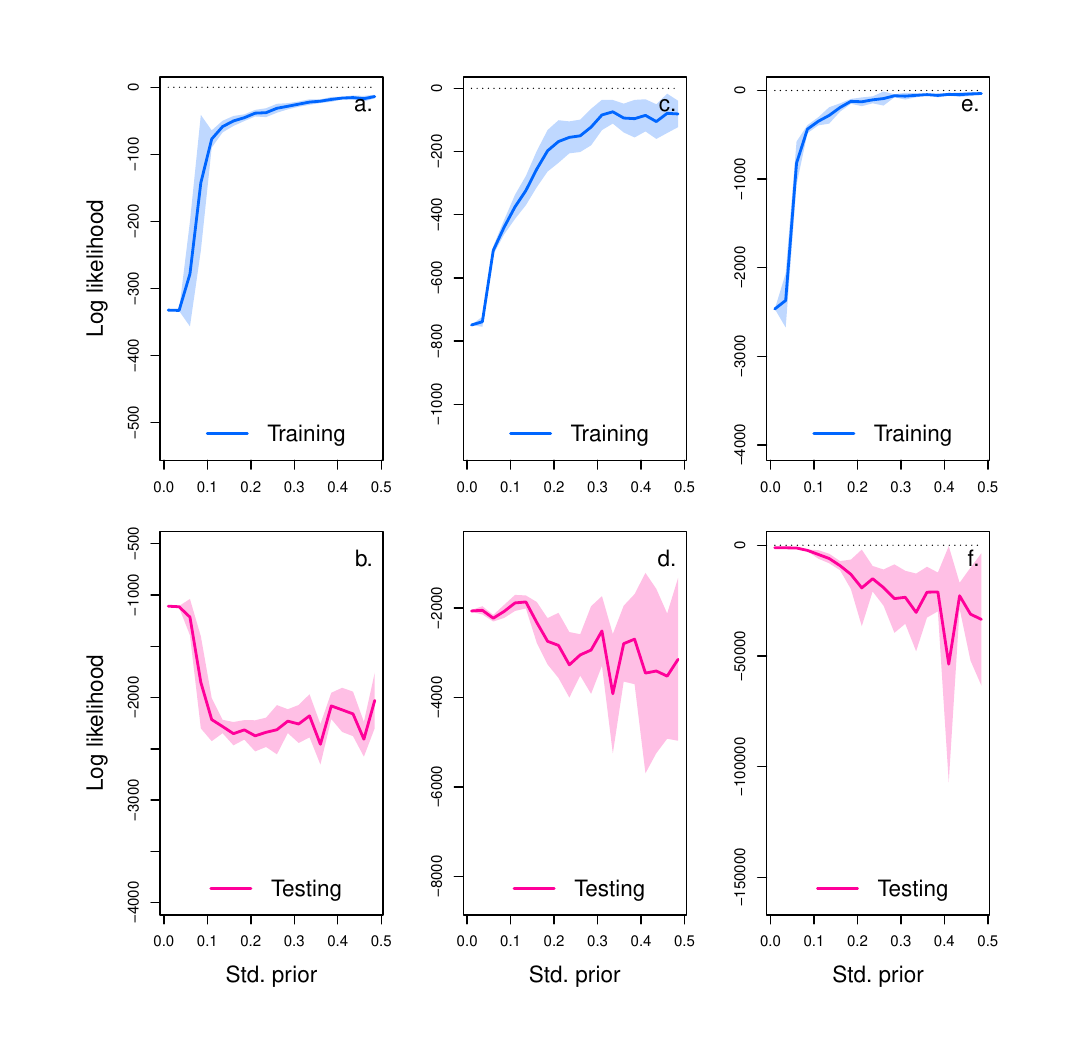}
\caption{
\textbf{Cross-validation plot of the NODE analysis of replicate B.}
The x-axis of the graphs correspond to the standard deviation of the prior distribution of the NODE parameters, which constrains the nonlinearity of the nonparametric approximation of the NODEs.
Small values of standard deviation correspond to a linear model, while higher values (towards 0.5) correspond to a highly nonlinear model.
Time series of algae, flagellate, and rotifer are split in half to create a train set and a test set. 
The model is fitted to the train set for increasing value of standard deviation (from 0.05 to 0.5 by 0.05 increments), and evaluted on the test set.
Graph a., c., and e. show the log likelihood of the NODE system fitted by BNGM to the train set of algae, flagellate, and rotifer, respectively.
Graph b., d., and f. show the log likelihood of the fitted NODE, evaluated on the corresponding test set.
The shaded areas represent the 90\% confidence interval on estimates, obtained by anchored ensembling of the log marginal posterior distribution (Eq. 7) (\cite{Pearce2018}).
}
\end{figure}
\newpage

%% figure
\newpage
\begin{figure}[H]
\includegraphics[width=1\linewidth,page=12]{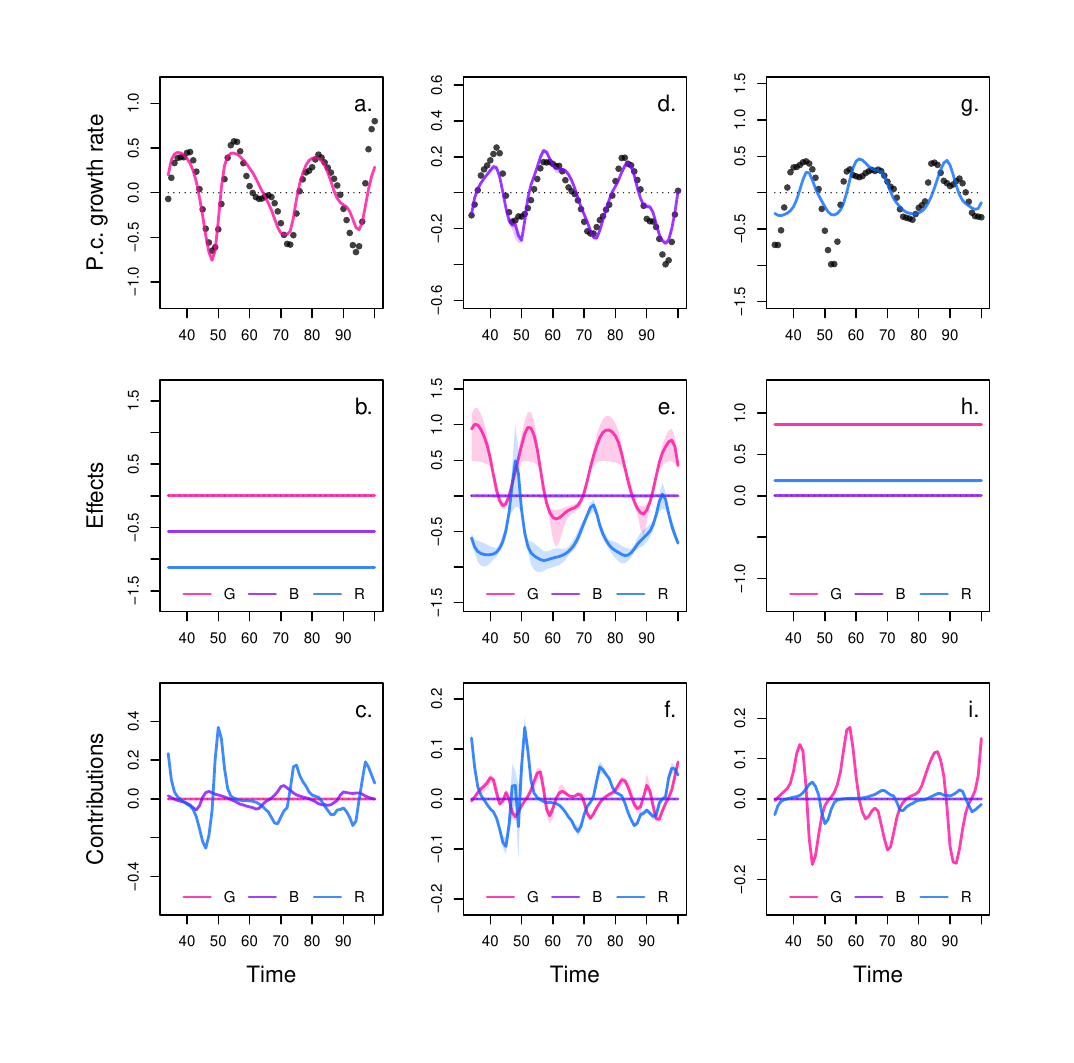}
\caption{
    \textbf{Drivers of dynamics of algae, flagellate, and rotifer in replicate B.}
    This figure displays the NODE nonparametric approximations of the per-capita growth rate of algae (a., b., c.), flagellate (d., e., f.), and rotifer (g., h., i.).
    We obtain the NODE approximations (a., d., g., solid line) by fitting the interpolated per-capita growth rates (black dots) with ANNs that take population densities as input.
    We then estimate the direction of ecological interactions (effects, b., e., h.) by computing the derivative of the NODE approximations with respect to each density.
    Finally, we compute the strength of ecological interactions (contributions, c., f., i.) by multiplying the interpolated dynamics of each population with its effects.
    The shaded area shows the 90\% confidence interval, obtained by approximately sampling the posterior distributions. 
    The replicated time series were obtained by digitising the time series in Hiltunen et al. (2013).
}
\end{figure}
\newpage

%%
% \newpage
\section{Complementary results case study 2 replicate C}

%% figure
% \newpage
\begin{figure}[H]
\includegraphics[width=1\linewidth,page=13]{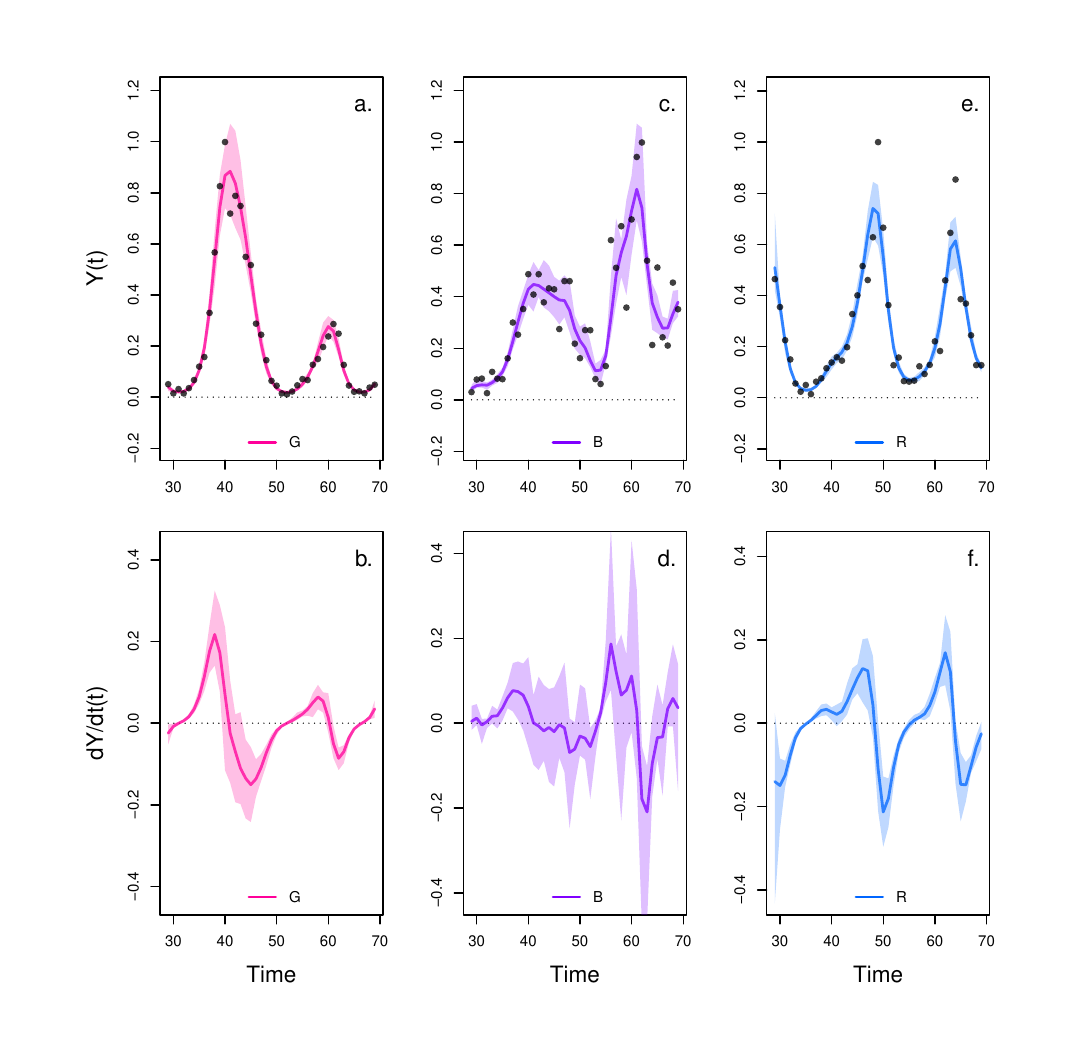}
\caption{
    \textbf{Interpolation of state and dynamics of algae, flagellate, and rotifer density in replicate B.}
    Graph a., c., and e. display the neural interpolation of the population density of algae (G), flagellate (B), and rotifer (R), respectively (obatined with Eq. 7). 
    Graph b., d., and f. show the corresponding interpolated dynamics, obtained by differentiating the interpolation of the states with respect to time (Eq. 5).
    The shaded areas represent the 90\% confidence interval on estimates, obtained by anchored ensembling of the log marginal posterior distribution (Eq. 7) (\cite{Pearce2018}).
    Time series are obtained from digitising the time series in \cite{Hiltunen2013}.
}
\end{figure}
\newpage

%% figure
\newpage
\begin{figure}[H]
\includegraphics[width=1\linewidth,page=14]{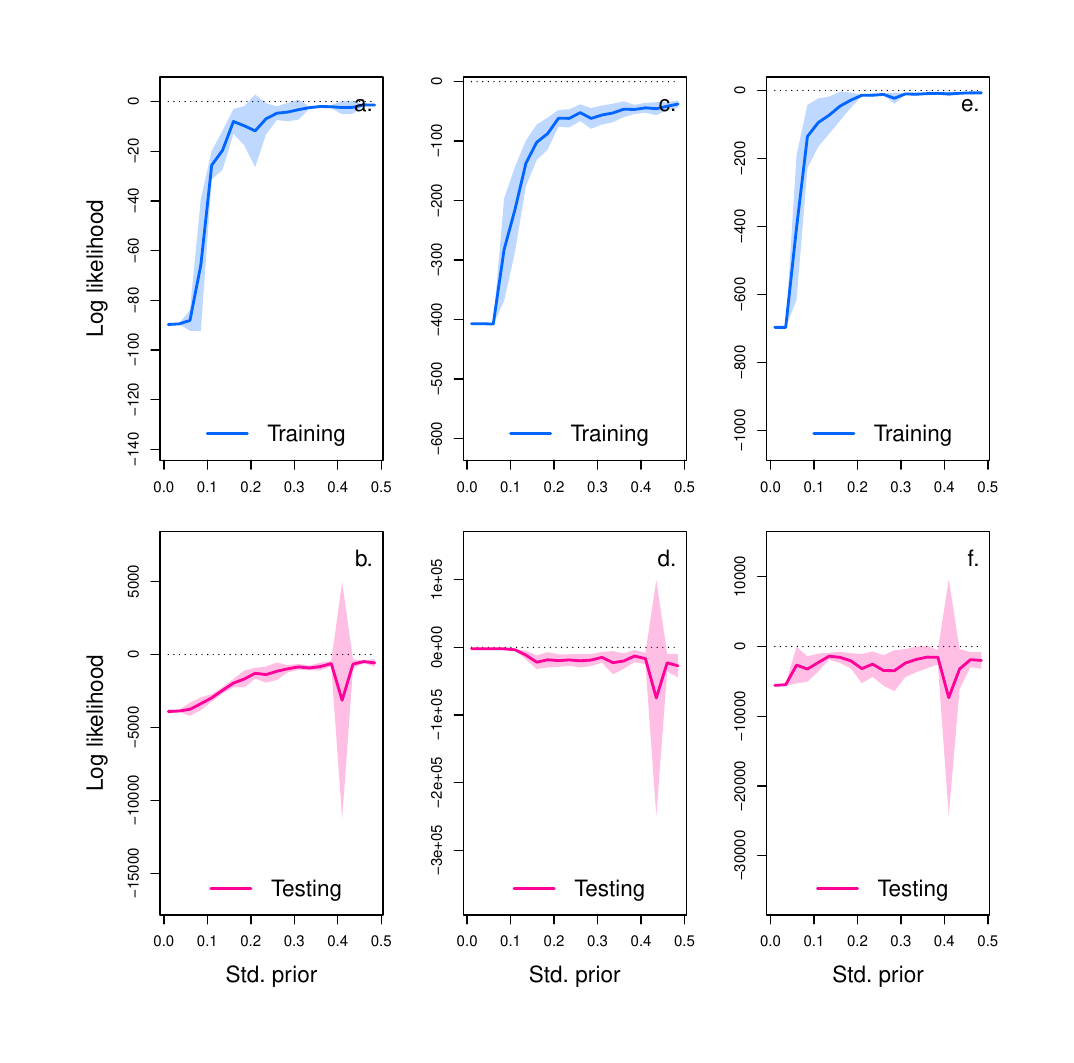}
\caption{
    \textbf{Cross-validation plot of the NODE analysis of replicate C.}
    The x-axis of the graphs correspond to the standard deviation of the prior distribution of the NODE parameters, which constrains the nonlinearity of the nonparametric approximation of the NODEs.
    Small values of standard deviation correspond to a linear model, while higher values (towards 0.5) correspond to a highly nonlinear model.
    Time series of algae, flagellate, and rotifer are split in half to create a train set and a test set. 
    The model is fitted to the train set for increasing value of standard deviation (from 0.05 to 0.5 by 0.05 increments), and evaluted on the test set.
    Graph a., c., and e. show the log likelihood of the NODE system fitted by BNGM to the train set of algae, flagellate, and rotifer, respectively.
    Graph b., d., and f. show the log likelihood of the fitted NODE, evaluated on the corresponding test set.
    The shaded areas represent the 90\% confidence interval on estimates, obtained by anchored ensembling of the log marginal posterior distribution (Eq. 7) (\cite{Pearce2018}).
}
\end{figure}
\newpage

%% figure
\newpage
\begin{figure}[H]
\includegraphics[width=1\linewidth,page=15]{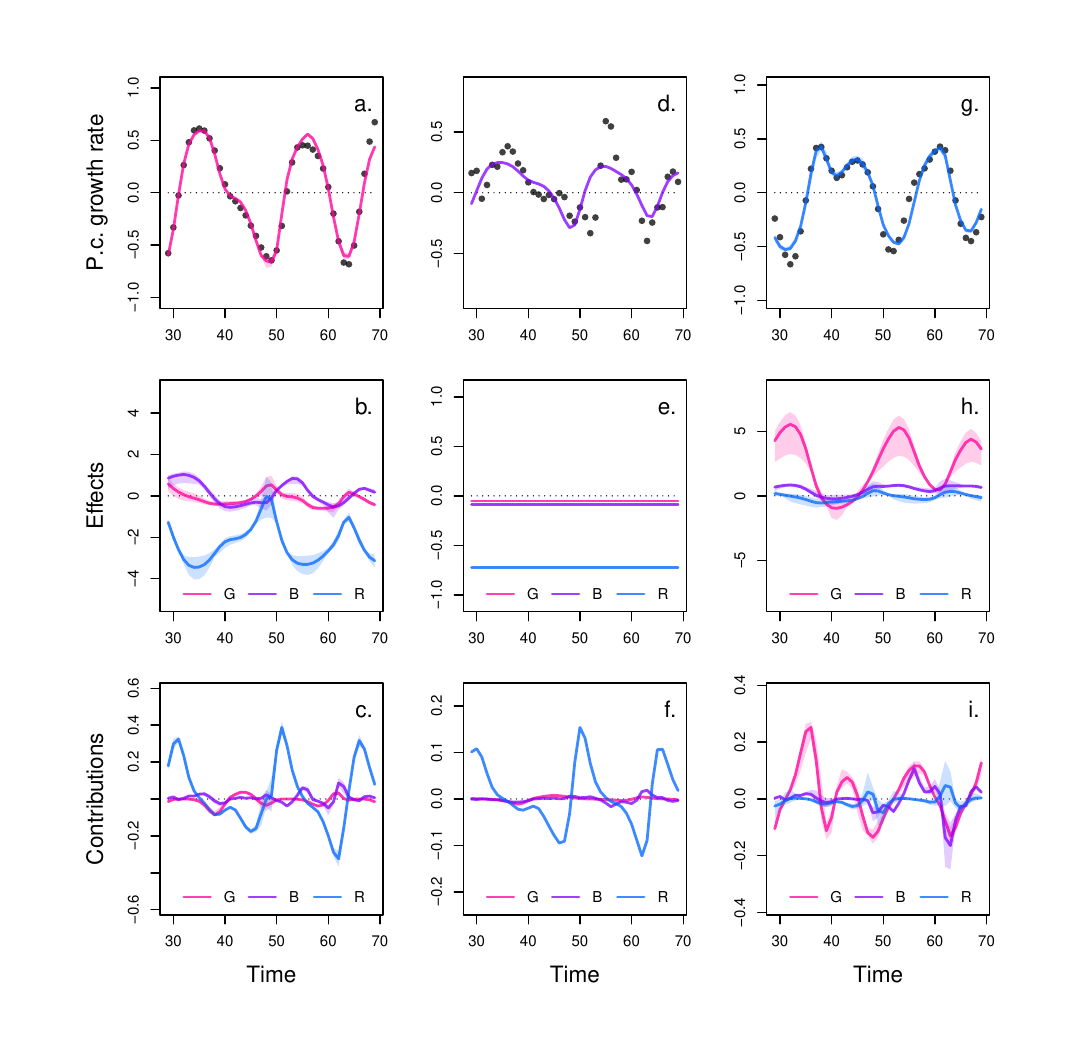}
\caption{
    \textbf{Drivers of dynamics of algae, flagellate, and rotifer in replicate C.}
    This figure displays the NODE nonparametric approximations of the per-capita growth rate of algae (a., b., c.), flagellate (d., e., f.), and rotifer (g., h., i.).
    We obtain the NODE approximations (a., d., g., solid line) by fitting the interpolated per-capita growth rates (black dots) with ANNs that take population densities as input.
    We then estimate the direction of ecological interactions (effects, b., e., h.) by computing the derivative of the NODE approximations with respect to each density.
    Finally, we compute the strength of ecological interactions (contributions, c., f., i.) by multiplying the interpolated dynamics of each population with its effects.
    The shaded area shows the 90\% confidence interval, obtained by approximately sampling the posterior distributions. 
    The replicated time series were obtained by digitising the time series in Hiltunen et al. (2013).
}
\end{figure}
\newpage

%%
% \newpage
\section{Complementary results case study 3}

%% figure
% \newpage
\begin{figure}[H]
\includegraphics[width=1\linewidth,page=16]{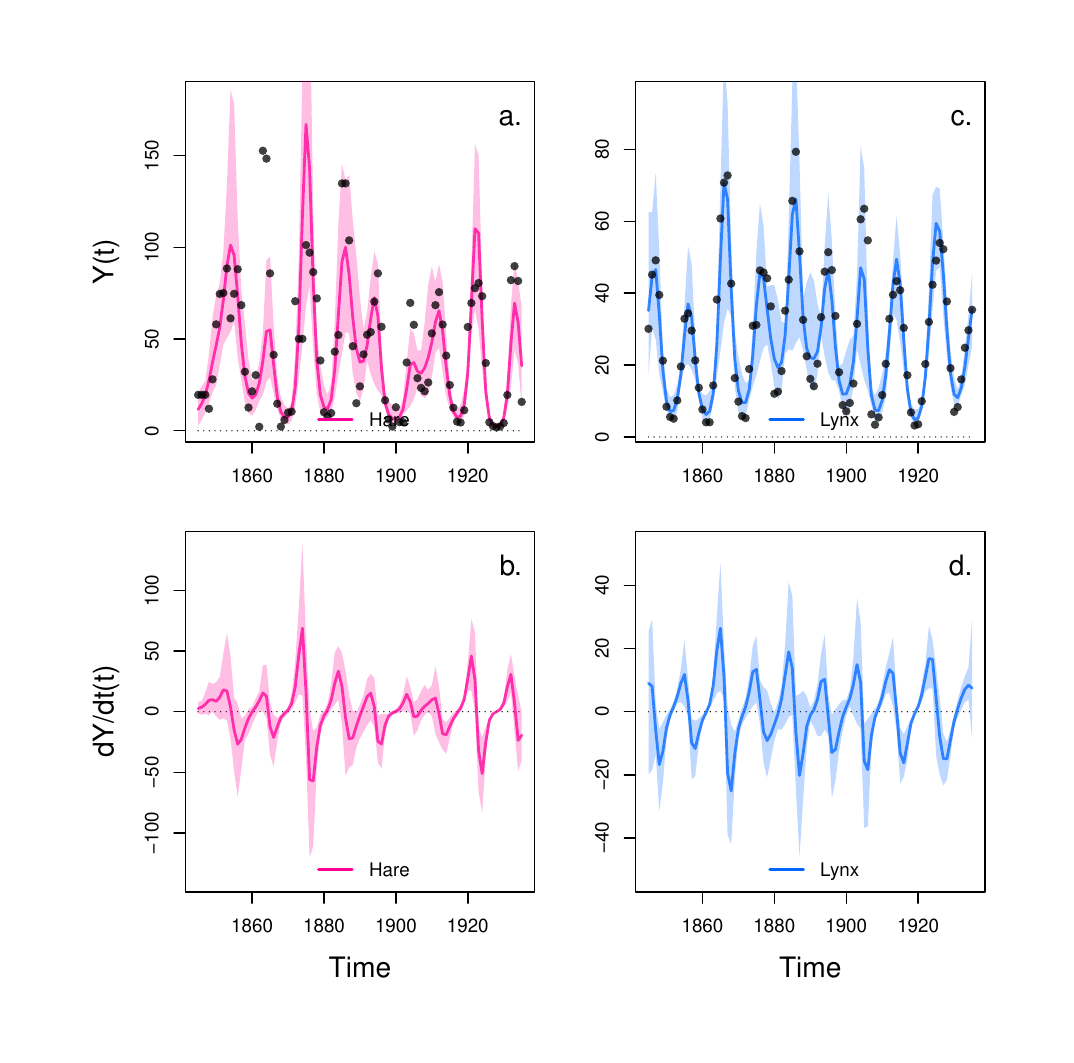}
\caption{
    \textbf{Interpolation of state and dynamics of hare and lynx.}
    Graph a. and c. display the neural interpolation of the population density of hare and lynx respectively (obatined with Eq. 7). 
    Graph b. and d. show the corresponding interpolated dynamics, obtained by differentiating the interpolation of the states with respect to time (Eq. 5).
    The shaded areas represent the 90\% confidence interval on estimates, obtained by anchored ensembling of the log marginal posterior distribution (Eq. 7) (\cite{Pearce2018}).
    Time series are obtained from \cite{Bonnaffe2021a}, originally from \cite{Odum1972}.
}
\end{figure}
\newpage

%% figure
\newpage
\begin{figure}[H]
\includegraphics[width=1\linewidth,page=17]{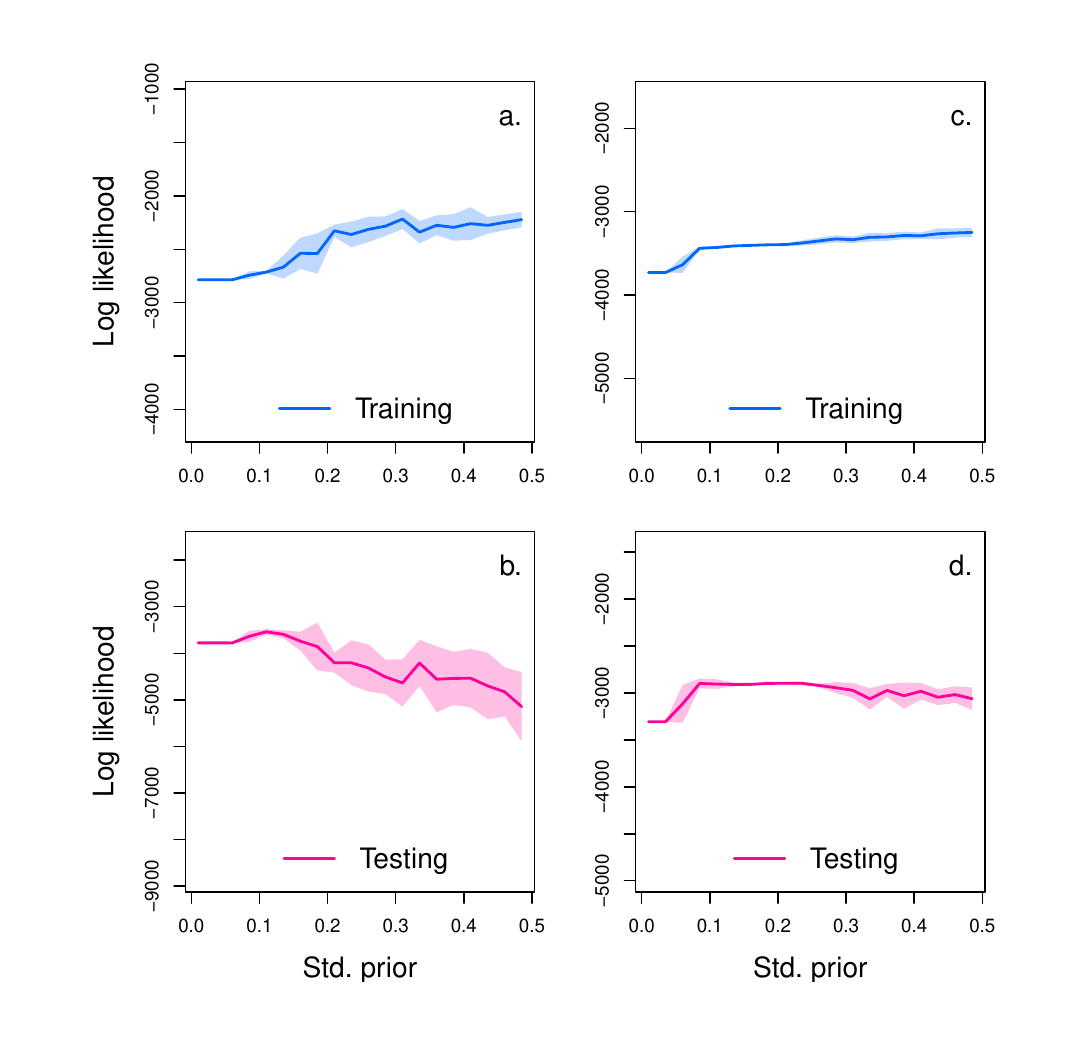}
\caption{
    \textbf{Cross-validation plot of the NODE analysis of the hare-lynx system.}
    The x-axis of the graphs correspond to the standard deviation of the prior distribution of the NODE parameters, which constrains the nonlinearity of the nonparametric approximation of the NODEs.
    Small values of standard deviation correspond to a linear model, while higher values (towards 0.5) correspond to a highly nonlinear model.
    Time series of algae, flagellate, and rotifer are split in half to create a train set and a test set. 
    The model is fitted to the train set for increasing value of standard deviation (from 0.05 to 0.5 by 0.05 increments), and evaluted on the test set.
    Graph a., c., and e. show the log likelihood of the NODE system fitted by BNGM to the train set of algae, flagellate, and rotifer, respectively.
    Graph b., d., and f. show the log likelihood of the fitted NODE, evaluated on the corresponding test set.
    The shaded areas represent the 90\% confidence interval on estimates, obtained by anchored ensembling of the log marginal posterior distribution (Eq. 7) (\cite{Pearce2018}).
}
\end{figure}
\newpage

%% figure
\newpage
\begin{figure}[H]
\includegraphics[width=1\linewidth,page=18]{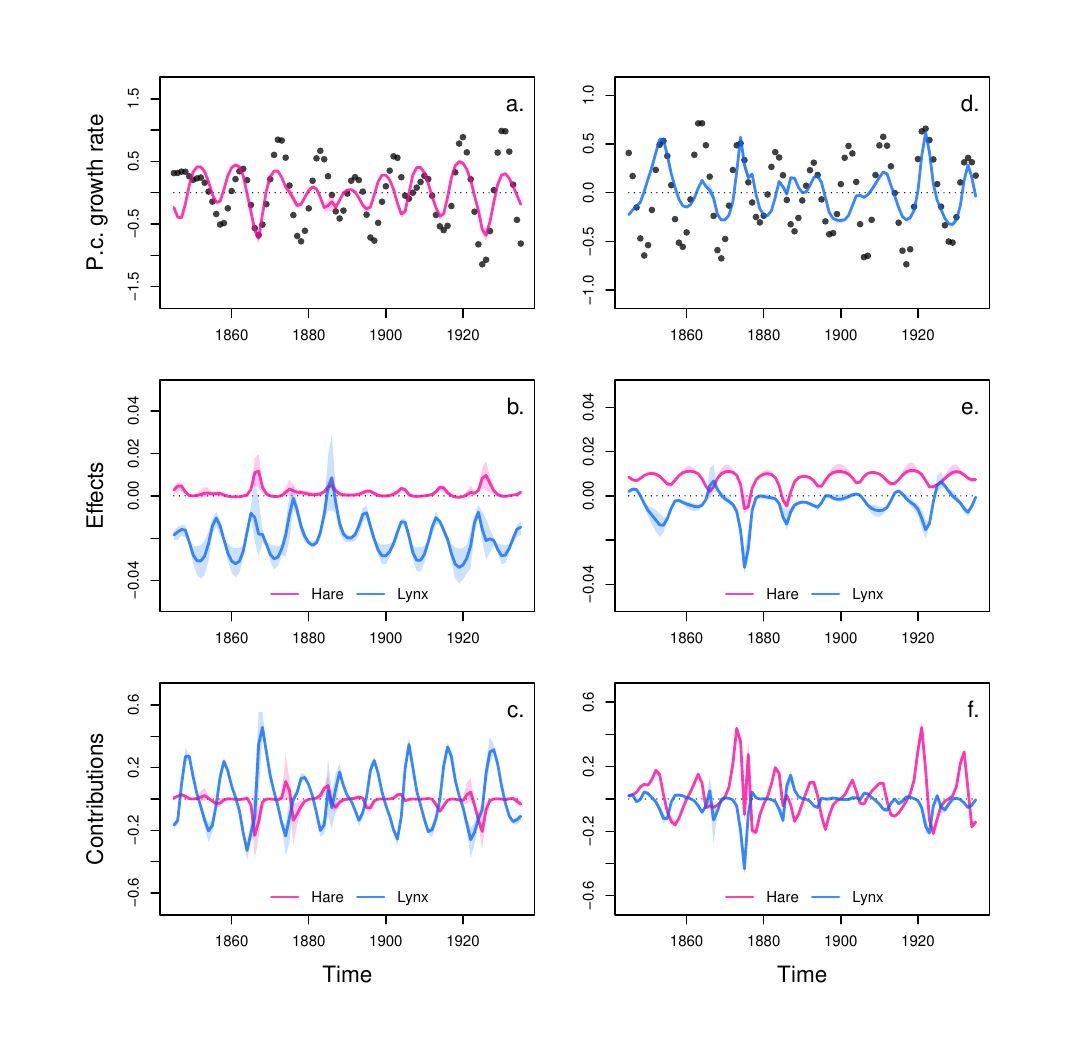}
\caption{
    \textbf{Drivers of dynamics of hare and lynx in the Odum and Barrett pelt count time series.}
    This figure displays the NODE nonparametric approximations of the per-capita growth rate of hare (a., b., c.), and lynx (d., e., f.).
    We obtain the NODE approximations (a., d., solid line) by fitting the interpolated per-capita growth rates (black dots) with ANNs that take population densities as input.
    We then estimate the direction of ecological interactions (effects, b., e.) by computing the derivative of the NODE approximations with respect to each density.
    Finally, we compute the strength of ecological interactions (contributions, c., f.) by multiplying the interpolated dynamics of each population with its effects.
    The shaded area shows the 90\% confidence interval, obtained by approximately sampling the posterior distributions. 
}
\end{figure}
\newpage

\end{document}